\documentclass[12pt]{article}
\usepackage{geometry}
\usepackage{graphicx}
\usepackage{amsmath}
\usepackage{amssymb}
\usepackage{xcolor}
\usepackage{stix} 
\usepackage{multibib}
\usepackage{authblk}
\usepackage{setspace}
\usepackage{outlines}
\usepackage{enumitem}
\usepackage{array}
\usepackage{etoolbox}

\AtBeginEnvironment{tabular}{\small}

\setenumerate[1]{label=\Roman*.}
\setenumerate[2]{label=\Alph*.}
\setenumerate[3]{label=\roman*.}
\setenumerate[4]{label=\alph*.}

\author[1,*,$\dagger$]{Timoth{\'e}e Poisot}
\author[1,*]{Marie-Andr{\'e}e Ouellet}
\author[2,3]{Nardus Mollentze}
\author[4]{Maxwell J. Farrell}
\author[5]{Daniel J. Becker}
\author[6]{Liam Brierley}
\author[7]{Gregory F. Albery}
\author[8,9]{Rory J. Gibb}
\author[10]{Stephanie N. Seifert}
\author[11,12]{Colin J. Carlson}

\affil[1]{D{\'e}partement de Sciences Biologiques, Universit{\'e} de Montr{\'e}al.}
\affil[2]{Institute of Biodiversity, Animal Health \& Comparative Medicine, University of Glasgow, Glasgow, UK}
\affil[3]{MRC - University of Glasgow Centre for Virus Research, Glasgow, UK}
\affil[4]{Department of Ecology and Evolutionary Biology, University of Toronto, Toronto, ON, Canada}
\affil[5]{Department of Biology, University of Oklahoma, Norman, OK, U.S.A.}
\affil[6]{Department of Health Data Science, University of Liverpool, Liverpool, UK.}
\affil[7]{Department of Biology, Georgetown University, Washington, D.C., U.S.A.}
\affil[8]{Centre on Climate Change and Planetary Health, London School of Hygiene and Tropical Medicine, London, UK.}
\affil[9]{Centre for Mathematical Modelling of Infectious Diseases, London School of Hygiene and Tropical Medicine, London, UK.}
\affil[10]{Paul G. Allen School for Global Health, WSU, Pullman, WA, U.S.A.}
\affil[11]{Center for Global Health Science and Security, Georgetown University Medical Center, Washington, D.C., U.S.A.}
\affil[12]{Department of Microbiology and Immunology, Georgetown University Medical Center, Washington, D.C., U.S.A.}
\affil[$\dagger$]{Correspondence should be directed to `timothee.poisot@umontreal.ca`}
\affil[*]{These authors share lead author status.}

\clearpage

\begin{document}
\title{Network embedding unveils the hidden interactions in the mammalian virome}

\maketitle
\setstretch{1}

\clearpage

\begin{abstract}
\noindent At most 1-2\% of the global virome has been sampled to date. Recent work has shown that predicting which host-virus interactions are possible but undiscovered or unrealized is, fundamentally, a network science problem. Here, we develop a novel method that combines a coarse recommender system (Linear Filtering; LF) with an imputation algorithm based on low-rank graph embedding (Singular Value Decomposition; SVD) to infer host-virus associations. This combination of techniques results in informed initial guesses based on directly measurable network properties (density, degree distribution) that are refined through SVD (which is able to leverage emerging features). Using this method, we recovered highly plausible undiscovered interactions with a strong signal of viral coevolutionary history, and revealed a global hotspot of unusually unique but unsampled (or unrealized) host-virus interactions in the Amazon rainforest. We develop several tests for quantifying the bias and realism of these predictions, and show that the LF-SVD method is robust in each aspect. We finally show that graph embedding of the imputed network can be used to improve predictions of human infection from viral genome features, showing that the global structure of the mammal-virus network provides additional insights into human disease emergence.
\end{abstract}

\clearpage
\newpage


\noindent The novel coronavirus SARS-CoV-2 is only the newest of the thousands of mammalian viruses that might have the capacity to infect human hosts. Despite growing interest in viral ecology, data remains limiting, as most  of the global virome remain undocumented. Computational methods that can infer undiscovered associations in a partially-observed host-virus network can fill-in some of these gaps \cite{albery2021science}. At least 20\% to 40\% of host-parasite associations are estimated to be unrecorded in locally-collected, highly-complete datasets \cite{dallas2017predicting}; a much higher proportion are likely unrecorded in the high-sparsity datasets cataloguing the global virome. An even greater proportion of host-virus interactions may be biologically plausible (i.e., a virus might have the capacity to infect a host) but as-yet unrealized for lack of ecological opportunities. These are often the links with the greatest relevance to actionable science: at least 10,000 mammalian viruses likely have the unrealized capacity to infect human hosts \cite{carlson2019global}, while an even greater number could be shared thousands of times between mammals as they track shifting habitats in a changing climate \cite{carlson2020climate}.

Here, we propose a novel method for predicting unknown links in partially-sampled networks, and apply it to the largest database of host-virus associations currently available. The method is based on a combination of \textit{linear filtering}, which uses high-level network information to generate an initial guess as to the probability of an interaction, and \textit{singular value decomposition}, which uses the structure of a low-rank approximation (which has a better signal-to-noise ratio \cite{nejati2016denoising}) of the entire network to impute interactions that were presumed negatives. In combination, this method uses existing knowledge on the entire network, but also can be tuned in such a way that its adjacency matrix is approximated at a rank that maximizes the amount of information used for imputation. Importantly, this method relies only on network structure, and does not consider (or require) external information specific to the hosts and viruses involved. We used this method to predict host-virus associations that are either undetected, or are biologically plausible but possibly unrealized in the real world. We found that the imputed network carries a weaker signal of sampling biases, while preserving the real signal of coevolutionary history. Applying methods from community ecology, we found that the Amazon may be more of a global hotspot of undescribed host-virus associations than currently assumed. Finally, we applied graph embedding to the observed and imputed networks, and used these as predictive features to augment a previously-published model that predicts which viruses can infect humans based on summaries of viral genome composition \cite{mollentze2020identifying}, testing whether knowledge about the global dynamics of cross-species transmission are informative for the limited case of predicting human disease emergence.

\subsection*{Predicting the host-virus network}

The combined linear filtering and singular value decomposition (LF-SVD) model relies on four hyper-parameters describing the relative importance of network structure (LF: in-degree, out-degree, and connectance) and matrix rank used for approximation (SVD). After tuning of the hyper-parameters, the best model (which used initial values emphasising network connectance, and performed SVD at rank 12) achieved a ROC-AUC of 0.84 (ED Table \ref{fig:svd-auc}, ED Figure \ref{fig:bestrocauc}). Although analyses of ecological networks usually gravitate towards using degree-based (over connectance-based) models, this choice of best model is unsurprising. Assuming that the overwhelming majority of interactions are unsampled (which is supported by the observation that imputation increased the number of interactions by a factor of about 15), degree holds very little information besides sampling effort.

We applied four tests of whether model performance was undermined by biases in the partially-observed network, a common problem in predicting host-pathogen interactions. First, we tested the effect of passive sampling bias with a regression of host species' viral diversity against citation counts, a commonly-used proxy for scientific research effort; we found that consistently, citations had a weaker effect predicting viral richness after imputation (ED Table \ref{tab:citations}). Second, we examined the top ten hosts that shared viruses with humans, to assess the influence of impact bias, a specific form of active sampling bias driven by relevance to human health. We found that while domesticated or lab animals dominated this list in the observed network, the imputed network removed most of these species (ED Figure \ref{fig:human10share}). Third, we assessed phylogenetic patterns in the number of `missing' viruses linked to each host after imputation. Missing viruses displayed only moderate phylogenetic signal (Pagel's $\lambda=$ 0.35), suggesting they are distributed fairly equally across the mammalian tree of life---a finding that matches other recently-published observations \cite{mollentze2020viral}. An additional taxonomic analysis identified four clades--the cetaceans, a subclade of mostly insectivorous bats, and two subclades of New World rodents--with fewer missing viruses than other mammals (ED Figure \ref{fig:phylofactor}, ED Table \ref{tab:phylofactor}). This suggests sampling efforts in these taxa may be more complete than viral sampling more generally \cite{wille2021accurately}. Finally, we mapped the global geographic hotspots of interactions based on mammal ranges, and found significant under-representation in the Amazon and Congo basins; these coldspots are significantly reduced (though not fully eliminated) in the imputed network, suggesting that our model is able to successfully anticipate the thousands of undiscovered ecological interactions that occur in hyperdiverse but undersampled tropical rainforests (ED Figure \ref{fig:geogbias}). 

\subsection*{Emergent properties of the imputed host-virus network}

Compared to the 5,494 interactions recorded in our original mammal-virus dataset, our model predicted a total of 75,901 new interactions (ED Figure \ref{fig:dotty}). With a total of 81,395 interactions, the imputed network has a connectance of 0.09, which is well within the range of connectances for antagonistic bipartite networks \cite{williams2011biology}. The best scoring model has a false discovery rate of 9.3\%, meaning that it is potentially over-predicting about 7,060 interactions. The same model has a false omission rate of 23\%, which would suggest a number of undiscovered interactions of the order of $10^5$ for this dataset. This being said, these numbers should be interpreted within the context of data constraints: the initial dataset is biased towards extreme sparsity, and for this reason it is likely that the imputed network is less severely incomplete than the false omission rate would suggest.

We next examined the post-imputation network for meaningful biological signals. The ``evolutionary distance effect'' is often the best-supported signal in host-virus networks: closely-related hosts share both viruses (through coevolution) and microbiologically relevant traits (through identity by descent) which facilitate cross-species transmission, leading to a correlation between evolutionary distance and virome similarity \cite{albery2020predicting}. We tested this property in both the pre- and post-imputation networks by examining viral sharing, both pairwise among all hosts and between humans and other mammals. We found a strong and consistent phylogenetic distance effect in both viral sharing (whether two hosts share any viruses at all) and the total number of viruses shared, both pairwise among mammals and specifically with humans (ED Figure \ref{fig:phylohuman}); although imputation reduced the signal of these effects, all but one (binary viral sharing with humans) remained significant even after imputation (ED Tables \ref{tab:phylohuman1}, \ref{tab:phylohuman2}). These results suggest that the missing interactions identified by our model have a high biological plausibility, and that---even though we do not incorporate phylogeny or any other host traits into our analyses---the latent factors that structure the network are identified and successfully recapitulated by the model.

\begin{figure}[!ht]
    \centering
    \includegraphics[width = 12cm]{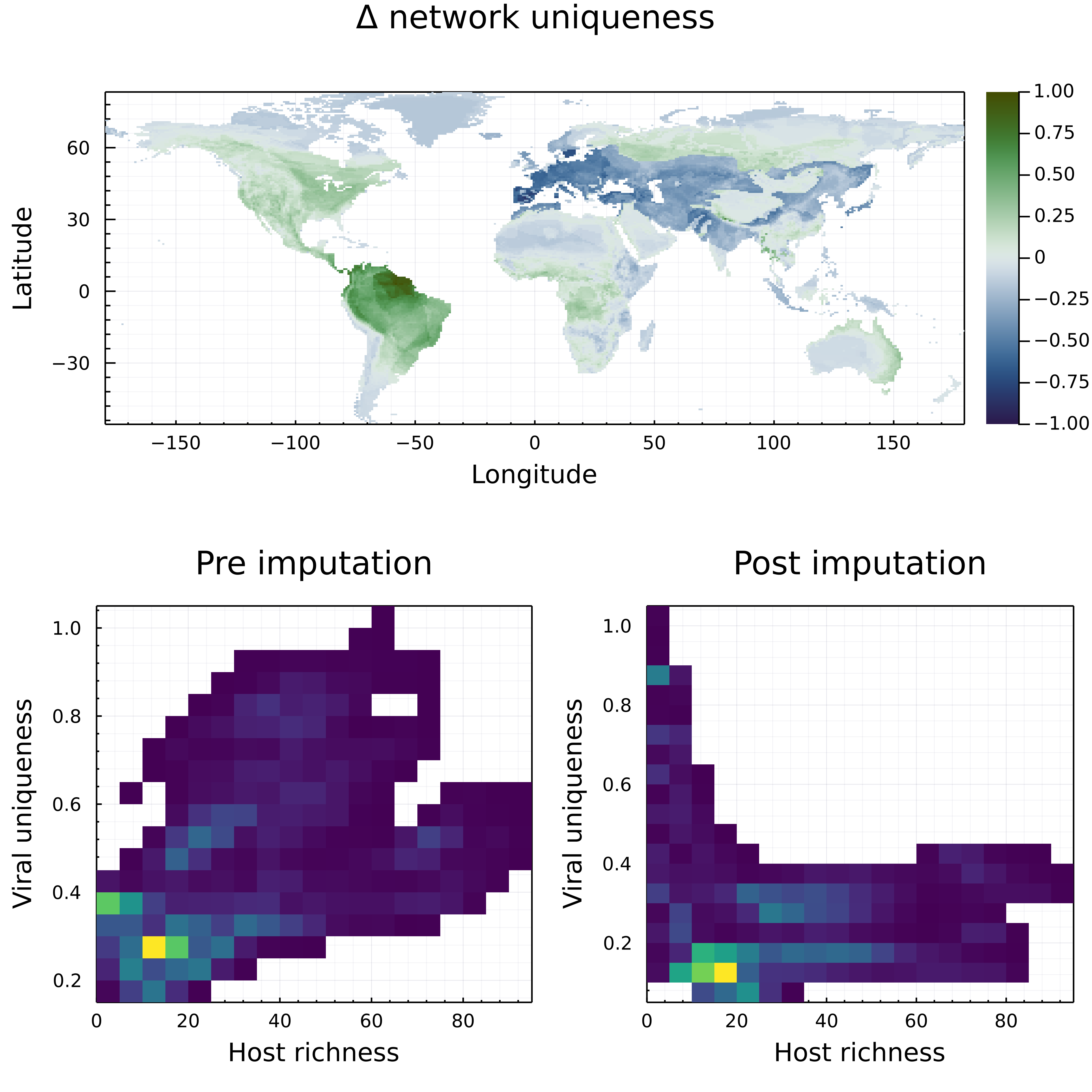}
    \caption{\textbf{Imputed network reveals undiscovered hotspots of unique host-virus associations in the Amazon}. Top: difference between (ranged) compositional uniqueness of the viral community based on host presence (see ED Fig. \ref{fig:lcbd-prepost}). Dark green areas indicate that the imputed network suggests a higher originality of the viral community than available data would. Bottom: comparison between the number of hosts and the uniqueness of the viral community. Assuming random discovery of viruses through host sampling, this relationship would be overall linear and positive, as is the case pre-imputation. Adding imputed interactions removes some of the sampling biases, and shows how areas with lower host richness have more unique contributions to viral uniqueness, which suggests that they harbor viruses not shared by more speciose locations.}
    \label{fig:lcbd}
\end{figure}

 Finally, we evaluated the effect of imputation on the spatial distribution of viral biodiversity. The \textit{local contribution to beta-diversity} approach~\cite{legendre2013beta}~-- essentially a partition of the variance in the community matrix -- measures the extent to which a single location differs from the expectation based on the entire range considered. When applied to interactions \cite{poisot2017hosts}, it reveals areas where, although the network might not be structurally different, it is composed of interactions that do not usually occur together. In biological terms, this means that novel host-jumps are possible through different host-virus pairs being in contact. Comparing the uniqueness of the viral community composition based on host spatial distribution before and after imputation reveals an undocumented hotspot of unique host-virus associations in the Amazon (Figure \ref{fig:lcbd}, ED Figure \ref{fig:lcbd-prepost}).

\subsection*{Predicting viruses with zoonotic potential}

We finally explored whether network-wide prediction offered useful insights into zoonotic potential, the ability of a virus to infect humans (a subset of links with one focal node in the network). Surprisingly, we found that the imputation method did not predict known human-associated viruses any better than random (AUC = 0.51; ED Table \ref{tab:svd-zoon}). This finding does reassuringly imply that zoonotic viruses are not contributing a particularly strong structural bias to the predictions, but indicates that the model performs poorly when predictions are restricted to one fairly-atypical node out of over 1,000. Indeed, while the ability of the model to predict the viruses associated with a given host generally increased as hosts are linked to more viruses, performance was poor for hosts linked to unusually-high numbers of viruses relative to the rest of the dataset (of which humans were the most extreme; ED Figure \ref{fig:connectivity}). A similar, but less extreme, pattern was observed among viruses linked to above-average numbers of hosts. Thus, although our best model focusing exclusively on connectance performed well in general, models incorporating in- or out-degree or specialized to a particular node may be needed for better-sampled nodes.

\begin{figure}[!ht]
    \centering
    \includegraphics[width = 15.5cm]{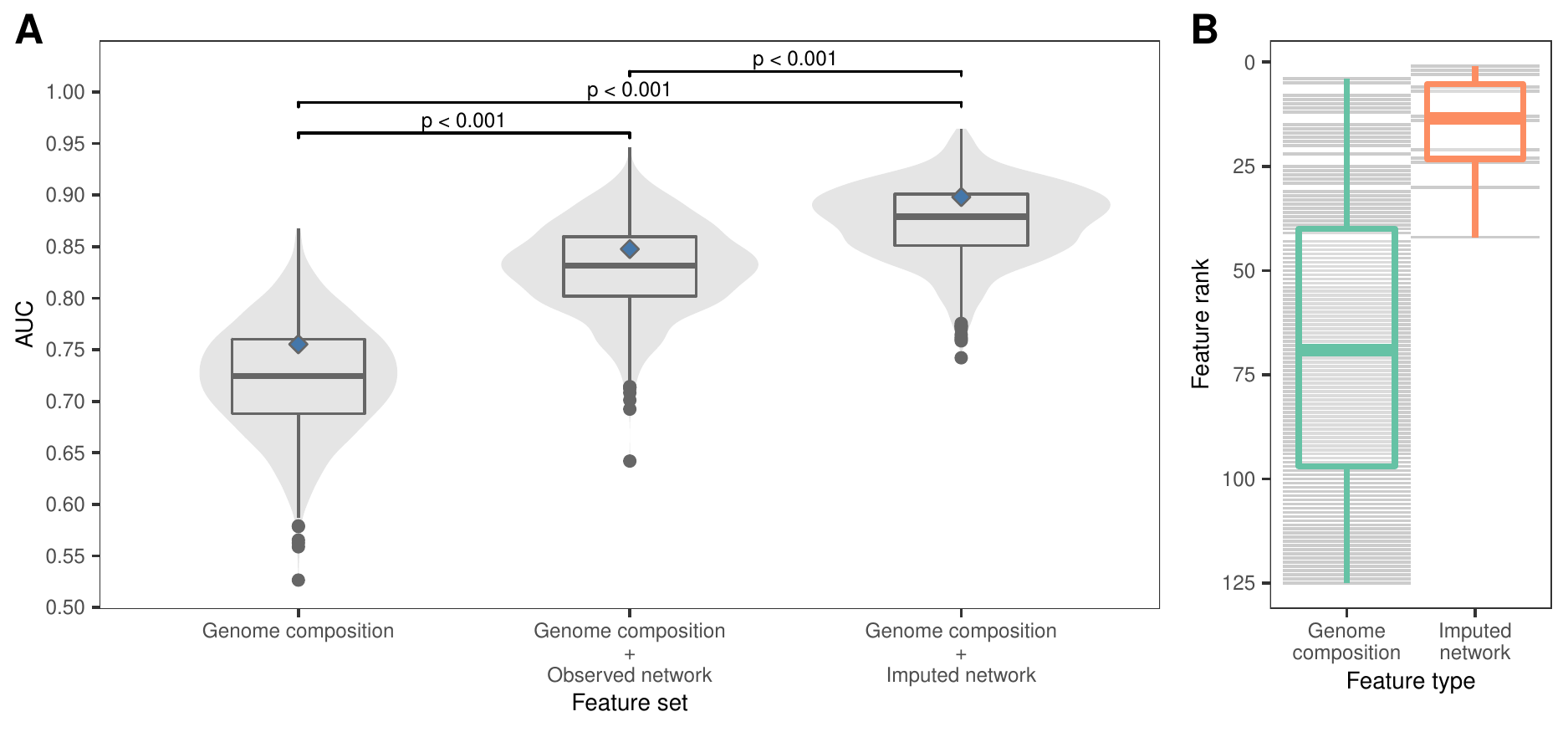}
    \caption{\textbf{Imputing the viral sharing network improved prediction of human infection ability.} (A) An existing model of human infection-risk using viral genomic features is improved when network embeddings are added as viral traits; models that use embeddings from the imputed network perform better than those using the observed network. Violins and boxplots show the ROC-AUC for test-set predictions across 1000 replicate 70\%:15\%:15\% train:calibrate:test splits ($N = 612$). P-values from pairwise Kruskall-Wallis rank sum tests are shown for all comparisons. Diamonds indicate the performance of a bagged model which averages predictions from the 100 best-performing models, based on test-set AUC iteratively re-calculated while excluding the virus being predicted. (Mean AUC: genome composition model $=$ 0.723; genome composition + observed network $=$ 0.830; genome composition + imputed network $=$ 0.875.) (B) Predictive feature importance in the combined (genome composition + imputed network) model; network embeddings are consistently the top predictive features, compared to biologically-informative measures of genome composition.}
    \label{fig:nardus}
\end{figure}

We next investigated whether the imputed host-virus network could be applied in specialised models aimed at identifying human-infecting viruses. Viral host breadth is a widely-used predictor of zoonotic ability, but is generally unavailable for poorly-studied viruses \cite{olival2017host, grange2021ranking, mollentze2020identifying}. To test whether the structural information on host range from our imputed network can be made accessible for prediction, we revisited a recently-developed model that applies boosted regression tree models to predict zoonotic potential based on the genome composition of animal viruses \cite{mollentze2020identifying}. We extracted the position of viruses in the pre- and post-imputation networks by removing humans (as well as viruses linked only to humans in the observed data) and applying random dot product graph embedding, which generated a total of 12 latent features that describe each virus's relationship to other viruses and animal hosts in the network. We then added these features to the genome composition-based model, and compared performance on the same set of viruses. Models incorporating the embeddings performed significantly better than a genome composition-only model, despite the fact that humans were removed from the network. Using embeddings derived from the post-imputation network consistently produced better predictions (mean test-set AUC = 0.875, SD = 0.04; Figure \ref{fig:nardus}). Averaging predictions across the top 10\% of repeated training iterations \cite{babayan2018predicting} further improved performance (AUC = 0.898). Moreover, of the top 20 viruses predicted by the algorithm, eleven already have serological or otherwise-circumstantial evidence of human infection (ED Table \ref{tab:nardus-zoon}), as do many of the other highly ranked viruses (ED Figure \ref{fig:serology}). The performance gains from using the imputed network suggest that LF-SVD is a viable option for improving applications reliant on knowledge of the global virome, including the identification of potential zoonotic viruses. However, more work is needed to establish the exact operating conditions under which such an approach can be safely applied; in particular, the number of animal hosts which need to have been found before reliable inferences on zoonotic risk can be made for novel viruses (cf. ED Figure \ref{fig:connectivity}) is difficult to assess without detailed data on the order in which hosts are linked to viruses (expected to be nonrandom given sampling biases).

\subsection*{Conclusions}
These results indicate that the structure of the observed host-virus network contains meaningful information about the rules of cross-species transmission. The imputation process recovers more of this information, even without the use of mechanistic predictors like host phylogeny, retaining biologically relevant signals while reducing key biases in current observational data. Thus, future efforts to predict viral emergence may be able to leverage the use of recommender systems as a data-inflation step to make better predictions. However, these approaches (and notably their validation) remain limited by how poorly characterized the host range of most viruses is; the majority of viruses are either undiscovered or known from a single host. As the global virome becomes better sampled, these approaches will be increasingly reliable not just for biological inference but for actionable efforts to prevent zoonotic emergence.

\clearpage

\noindent\textbf{Acknowledgements}: This work was supported by funding to the Viral Emergence Research Initiative (VERENA) consortium including NSF BII 2021909 and a grant from Institut de Valorisation des Données (IVADO). NM was funded by the Wellcome Trust (217221/Z/19/Z). TP and MAO were funded by the Fondation Courtois. This research was enabled in part by support provided by Calcul Québec (www.calculquebec.ca) and Compute Canada (www.computecanada.ca). We acknowledge that this study was conducted on land within the traditional unceded territory of the Saint Lawrence Iroquoian, Anishinabewaki, Mohawk, Huron-Wendat, and Omàmiwininiwak nations.

\bigskip

\noindent\textbf{Code availability statement}: The code for LF-SVD tuning, imputation, and analysis, has been archived at \texttt{10.5281/zenodo.4850581}. The code for prediction of zoonotic potential has been archived at \texttt{10.5281/zenodo.4850643}.

\clearpage

\section*{Methods}

\subsection*{Model design and implementation}

\subsubsection*{Host-virus association data}

We used a recently published dataset called CLOVER \cite{gibb2021data}, which is the largest open dataset describing the mammal-virus network currently available, and combines data from four sources that each cover overlapping-but-distinct portions: the Host-Pathogen Phylogeny Project (HP3) dataset \cite{olival2017host}, the ENHanCEd Infectious Diseases Database (EID2) \cite{wardeh2015database}, the Global Mammal Parasite Database version 2.0 (GMPD2) \cite{stephens2017global}, and an unnamed dataset recently published by Shaw \textit{et al.} \cite{shaw2020phylogenetic}. By reconciling these datasets and their underlying taxonomy, the CLOVER dataset is able to achieve a 30\% reduction in matrix sparsity over the next most detailed dataset.

The CLOVER dataset describes 5,494 interactions between 829 viruses and 1,081 mammalian hosts. The majority of these interactions have been recorded in wild animals, using a combination of detection methods (usually serology, PCR, or viral isolation). A small portion of records assimilated from NCBI's GenBank into these other datasets may also record experimental infections, which provide insight into biological compatibility but not necessarily opportunity for infection in nature. Each of the component datasets, and the CLOVER dataset, are presence-only (i.e., they only report an edgelist of known interactions, and do not include true negatives).

\subsubsection*{Imputation model description}

The imputation model uses two steps to chain linear filtering (which can recommend potentially false-negative interactions \cite{stock2017linear}) to recommendation based on singular value decomposition (which adequately captures the low-rank structure of ecological association networks \cite{strydom2020svd}). This imputation model is hereafter termed LF-SVD. The LF step relies on four hyper-parameters expressed as an array of weights $\mathbf{\alpha}= [\alpha_1, \alpha_2, \alpha_3, \alpha_4]^T$, which are respectively the relative importance of the original (i.e., observed) value of the interaction, in- and out-degree, and of connectance  (the constraint $\sum \mathbf{\alpha} = 1$ is always enforced). LF creates a potential matrix $\mathbf{A}$ from an observed matrix $\mathbf{Y}$ of size $(n,m)$ by assigning every interaction between species $i$ and $j$ an initial score given by the dot product of weights and properties of $\mathbf{Y}$,

$$A_{ij} = [Y_{ij}, \frac{1}{n}\sum_k Y_{kj}, \frac{1}{m}\sum_l Y_{il}, \frac{1}{nm}\sum \mathbf{Y}]\cdot \alpha  \,.$$

\noindent This corresponds to a weighted average of averages, wherein $\forall (i,j), 0 \le A_{ij} \le 1$. We compared three parameterizations of the $\alpha$: connectance only ($[0, 0, 0, 1]^T$), degree only ($[0, 1, 1, 0]^T$), and hybrid ($[0, 1, 1, 1]^T$). While technically there is an infinite number of possible configurations for the LF weight vector, the computational cost of a grid search is prohibitive, and these parameterizations have the added benefit of corresponding to phenomenological assumptions about what drives network structure that have been well laid out in the literature \cite{williams2011biology}. In this application, we set $\alpha_1=0$, as the initial value of the interaction is ignored, reducing th number of hyper-parameters to tune from four to three. 

We updated the initial values produced by LF using (truncated) singular value decomposition (SVD) imputation. Like Principal Component Analysis (PCA), SVD is an embedding of a starting matrix into latent subspaces; compared to PCA, SVD is a more general solution that also handles numerical instability due to very small entries well \cite{Shlens2014TutPri}, which is a likely scenario as some interaction probabilities are expected to be small. As all entries of both $\mathbf{A}$ and $\mathbf{Y}$ are in $\mathbb{R}$, we can decompose either of these matrices as $\mathbf{U}\mathbf{\Sigma}\mathbf{V}^T$, where $\mathbf{U}$ and $\mathbf{V}$ are unitary matrices known as the left and right subspaces, and $\mathbf{\Sigma}$ is a diagonal matrix containing the singular vales of the decomposed matrix. To impute the interaction $(i,j)$, we create a matrix $\mathbf{K} = \mathbf{Y}$, wherein $K_{ij} = A_{ij}$ (according to the LF model). To decompose this matrix at low-rank $k$, we set the values of $\mathbf{\Sigma}$ larger than $k$ to 0, and calculate the approximate version of $\mathbf{K}$ as

$$\mathbf{\bar{K}} = \mathbf{U}_k\mathbf{\Sigma}_k\mathbf{V}_k^T$$

\noindent The overall SVD step was conducted as follows: for every interaction $(i,j)$,  we first set its value according to the LF model, and perform the truncated SVD step as outlined above. We then update $\mathbf{K}$ so that $K_{ij} = \bar{K}_{ij}$. The SVD step is repeated 20 times (after preliminary assays revealed that the absolute change after 10 iterations was consistently smaller than $10^{-3}$), and the final value after 20 iterations is the \textit{score} for the imputed interaction. Note that due to the nature of SVD, the score is not bound to $[0,1]$. The tuning and imputation of the LF-SVD model were performed in the software Julia 1.6 \cite{bezanson2012julia}, using the \texttt{EcologicalNetworks.jl} package \cite{poisot2019ecologicalnetworks}. 

\subsubsection*{Hyper-parameters tuning, thresholding, and evidence scoring}

In order to tune the hyper-parameters (LF weight vector, SVD rank), we picked a calibration set of 800 positives and 800 assumed negative interactions, and imputed them using each possible model ($n = 60$). This makes the strong assumption that the 800 negative interactions we picked in the calibration set were indeed true negatives; although the model ended up recommending a large number of interactions, ecological networks are known for their sparsity, and we judged this assumption acceptable based on an overall examination of model performance.

For each set of 1600 predictions returned by the models, we derived confusion tables at thresholds ranging from the lowest to the highest score, using a stepsize of 1000. From this confusion table, we calculated the ROC-AUC, true/false positive/negative rates, positive/negative predictive values, false discovery/omission rates, critical success index, accuracy, and informedness (\textit{a.k.a} Youden's J). The model with the highest ROC-AUC was picked as the best model, and used for the rest of this manuscript.

The exact cutoff to use to transform the continuous output of LF-SVD into a binary classifier (\textit{i.e.} the interaction is recommended or not) was determined by picking the threshold value maximizing Youden's J statistic. Each interaction is given an evidence score, which is obtained by dividing the values post-imputation (LF-SVD) by the values pre-imputation (LF), minus one. An evidence of 0 means that the imputation did not change the value, and increasingly positive values meant that the change due to imputation was stronger. This interaction evidence was used to rank interactions when required for the analyses.

\subsection*{Analysis of the imputed network}

\subsubsection*{Additional data sources}

For phylogenetic analyses, we used a recently published mammalian supertree published by Upham \textit{et al.} \cite{upham2019inferring}, which has been taxonomically harmonized to the CLOVER dataset for ease of analysis. For geographic analyses, we used the IUCN Red List (iucnredlist.org) species distribution maps for mammals, downloaded on June 6, 2019. For citation counts, we extracted total virus-related publications for each species (by searching for host species binomial plus all known synonyms and ``virus'' or ``viral'') from the PubMed database using the R package \texttt{rentrez} \cite{winter2017rentrez}.

\subsubsection*{Testing effects of biased data collection}

Observed host-pathogen association networks compiled from published records are influenced by a \textit{passive sampling bias} resulting from differential research across host and pathogen species. In comparative analyses of viral richness per host species, the number of publications per host species is often included as a covariate in an attempt to control for variable sampling effort across hosts \cite{teitelbaum2020estimators}. This estimate of sampling bias is consistently positively related to viral richness, and typically is the strongest predictor, explaining more variation than other biological covariates \cite{olival2017host,nunn2003comparative,ezenwa2006host,huang2015parasite,luis2013comparison}. To explore whether network imputation via LF-SVD is extrapolating sampling biases across host species, we conducted a set of phylogenetic regressions of the relationship between viral richness and the number of publications per host species (both in total and limited to those including keywords about viruses). Models were fit using the formulation of phylogenetic least squares regression provided via the \texttt{pgls} function (Pagel's lambda estimated via maximum likelihood) in the R package \texttt{caper} \cite{freckleton2002phylogenetic, orme2012caper}. By comparing models of observed viral richness to estimates after imputation with LF-SVD, we investigate the slope of the relationship and the explained variance in viral richness to assess how strongly passive sampling biases are retained in the LF-SVD imputed network. 

In addition to passive sampling bias, host-virus association data are frequently shaped by \textit{active} or \textit{impact} bias, where surveillance is targeted based on relevance to human health or economics. This is easily detected in records of viral sharing with humans. In principle, the species with the highest similarity to the human virome should be species that are closely related to humans (primates) or frequently live alongside humans (domesticated animals or synanthropic wildlife, particularly rodents that can live in human settlements), but domesticated animals and laboratory model systems will also score disproportionately in this metric, because of sampling effort. As a new test of model bias, we propose that imputation should reduce the signal of the latter group in viral sharing with \textit{Homo sapiens}, leaving mostly the former. To test the effect of active sampling bias, we examined the top 10 hosts based on similarity to \textit{Homo sapiens}, pre- and post-imputation. Prior to imputation, the top 10 list (based on Jaccard similarity of host and human viral community) includes six livestock or companion animals (\textit{Bos taurus}, \textit{Equus cabalus}, \textit{Sus scrofa}, \textit{Ovis aries}, \textit{Capra hircus}, and \textit{Canis lupus familiaris}), three primates (\textit{Pan troglodytes}, \textit{Macaca mulatta}, and \textit{Macaca fascicularis}), and one synanthropic and commonly-studied laboratory animal (\textit{Mus musculus}). After imputation, four of the domesticated or primate species remained (\textit{Canis lupus familiaris}, \textit{Equus caballus}, \textit{Sus scrofa}, and and \textit{Pan troglodytes}). The updated list includes two more primates (\textit{Gorilla beringei}, \textit{Gorilla gorilla}) and four more mice or rats (\textit{Hylaeamys megacephalus}, \textit{Permoyscus maniculatus}, \textit{Proechimys guyannensis}, and \textit{Zygodontomys brevicauda}). This mostly reflects changes in the network connectivity; all but one of these are in the top 10 species to gain links (with \textit{Z. brevicauda} replaced by \textit{Rattus rattus}).

\subsubsection*{Phylogeographic signals of missing interactions}

The distribution of missing viruses (each host species' total number of predicted but unknown host-virus links) across space, and across the evolutionary tree, are interlinked patterns that are of significant interest to viral ecologists \cite{olival2017host}. These patterns inform scientists' understanding of where undiscovered zoonotic threats might emerge and can be used to target sampling to locations and taxa with the greatest number of undiscovered viruses. However, these predictions are also difficult to disentangle from sampling bias, which can create spurious patterns that are undermined on closer analysis \cite{mollentze2020viral}. 

To assess phylogenetic patterns in the number of missing viruses, we used the previously-specified supertree \cite{upham2019inferring}. To match virus data against the phylogeny, we averaged missing virus counts for 30 species (n = 14 tips in the supertree). We used the \texttt{caper} R package to first broadly estimate phylogenetic signal as Pagel's $\lambda$ \cite{pagel1999inferring}. We next applied a graph partitioning algorithm, phylogenetic factorization, to more flexibly identify mammal clades that differ in missing virus counts. We used the \texttt{phylofactor} R package to partition counts of missing viruses in a series of generalized linear models with a negative binomial distribution \cite{washburne2019phylofactorization}. We determined the number of significant clades using Holm's sequentially rejective test with a 5\% family‐wise error rate.

We identified weak-to-moderate overall phylogenetic signal in the number of missing viruses ($\lambda$ = 0.35), although this estimate was distinct from both phylogenetically indepent models and Brownian motion models of evolution (both p < 0.01). Phylogenetic factorization in turn identified only four small clades with significantly different counts of missing viruses, all of which had fewer missing viruses than the remaining mammal phylogeny (ED Table \ref{tab:phylofactor}). These clades included ceteceans ($\bar{x}$ = 18, n = 30) and a subclade of primarily insectivorous Yangochiroptera ($\bar{x}$ = 43, n = 109) as well as two subclades of the New World rodent subfamily Sigmodontinae ($\bar{x}$ = 11, n = 11; $\bar{x}$ = 16, n = 15). Overall, these results indicate that, with the exception of some coldspots likely driven by oversampling (or, in the case of cetaceans, a peripheral role in the host-virus network), missing viruses are distributed fairly equally across the mammalian tree of life---a finding that matches other recently-published work \cite{mollentze2020viral}.

To assess geographic patterns in the number of missing viruses, we evaluated the number of known and missing viruses at the level of each host species and joined these to each host's IUCN range map. We mapped the total number of hosts with recorded interactions, the total number of known and predicted missing interactions, and the normalized difference between missing interactions and host diversity. Known interactions are recorded disproportionately in Europe and Asia, and to a lesser degree North America, a pattern that reveals strong sampling bias in viral inventories (ED Figure \ref{fig:geogbias}). This pattern is substantially reduced in the missing interactions, which globally track the true distribution of mammal diversity fairly well (better, in some places, than the hosts with viral interactions recorded in CLOVER). However, the normalized difference map still revealed a bias towards interactions predicted in North America and Eurasia, with coldspots in South America and Africa (ED Figure \ref{fig:geogbias}).

\subsubsection*{Coevolutionary signal in viral sharing}

To test for the signal of evolutionary history in the viral sharing network, we analyzed two outcome variables (viral sharing as a binary state, and as the total number of viruses shared) for two data structures (the entire pairwise host-host viral sharing matrix, or each hosts' sharing with \textit{Homo sapiens}, i.e., its role in zoonotic disease) in both the pre- and post-imputation network. We analyzed these variables as a function of phylogenetic distance using generalized linear models (GLMs), with viral sharing coded as a binomial outcome (logit link) and the count data modeled using a Poisson distribution. GLMs were fit using the \texttt{stats} package in R, and adjusted R-squared values were derived using the \texttt{rsq} package. Model coefficients and significance are given in ED Tables \ref{tab:phylohuman1} and \ref{tab:phylohuman2}. Response curves were finally plotted using the automated smoothing in the \texttt{ggplot} package with the same specifications.

We found that viral sharing, as a binary outcome, decoupled substantially from phylogeny after imputation. In large part, this can be explained by the fact that, with a 16-fold increase in connectance, binary sharing should become substantially less informative after imputation. (This also makes biological sense: for example, nearly all mammal species should share the capacity to be infected with true generalist viruses like rabies and influenza A.) In particular, the phylogenetic signal of viral sharing with humans became insignificant (p = 0.52) after imputation, the only insignificant relationship among those we tested. While the count data also recovered a reduction in effect size after imputation, we found that this reduction was much smaller, and that the phylogenetic signal of sharing with \textit{Homo sapiens} was slightly more explanatory in the post-imputation network.

\subsubsection*{Community uniqueness analysis}

We performed a measure of community compositional uniqueness using the local contribution to beta-diversity approach \cite{legendre2013beta}, and specifically its extension to interaction data following \cite{poisot2017hosts}. LCBD identifies locations (here, pixels) in which the community composition contributes more to the overall dissimilarity. For this section, we will note $\mathbf{X}$ the sites-by-items matrix, often referred to as a ``community data matrix,'' in which locations are rows, and items (host,viruses, interactions) are columns. The total beta-diversity is measured as $\beta = \text{Var}(\mathbf{X})$, after rows and columns with a marginal sum of 0 have been removed. The $\mathbf{X}$ matrix is then transformed by centering and squaring the values, so that $\mathbf{S} = [S_{ij}] = [(X_{ij}-\bar{X_{j}})^2]$. The sum of squares in $\mathbf{X}$ is then simply given by $\text{SS}_\text{total} = \sum_i \sum_j S_{ij}$. From there, measuring the LCBD (\textit{i.e.} the actual contribution of each location to $\beta$) is done by summing the matrix $\mathbf{X}$ row-wise, and dividing by the total sum of squares:

$$\text{LCBD}_i = \sum_j \frac{S_{ij}}{\text{SS}_\text{total}}\,.$$
Within every location, this value indicates the degree of uniqueness of this location (sampling unit) compared to all other sampling units in the data. LCBD values are typically, but not necessarilly, measured after $\mathbf{X}$ has been transformed using Chord's or Hellinger's distance. This, however, assumes that sampling is close to complete, which is an unreasonable assumption in our observed dataset; as applying an Hellinger transformation post- but not pre-imputation would prevent a comparison of the results, we work on the raw matrices.

\subsubsection*{Prediction of zoonotic potential}

We next tested whether the expanded host-range information available in the imputed network could improve zoonotic risk prediction in cases where information on individual viruses is limited. We expanded a recently developed model which combines summary statistics of viral genome composition and compositional similarity to human genes to predict zoonotic risk \cite{mollentze2020identifying}.

We extracted pseudo-traits from the host-virus network using latent variables, by extracting the left latent subspace of a random dot product graph decomposition \cite{dalla2016exploring}. We used the same number of dimensions (12) as for the low-rank approximation based on the imputation method. The feature matrix for viruses is given by

$$\mathbf{\bar{F}} = \mathbf{U}_{12}\sqrt{\mathbf{\Sigma}_{12}}\,$$

\noindent where $\mathbf{U}$ and $\mathbf{\Sigma}$ are the truncated left-subspace and singular values matrices of the decomposition of the network. This method was selected because the latent traits extracted this way can reproduce the original network within an arbitrary precision threshold, and have been shown to capture evolutionary signal on network structure \cite{dalla2016exploring}. To avoid leaking data on observed human infection into subsequent model training and evaluation steps, these network embeddings were generated while excluding humans. We also removed all viruses which had thus far only been linked to humans, since -- after the removal of humans from the network -- these viruses were uniquely identifiable as some of the only included viruses with no links in the network (another potential data leak; a small number of viruses with no known mammalian hosts were similarly unlinked, but these were rare enough that a model which predicted all unlinked viruses as human-infecting would have had reasonably high performance). 

Full genomes were available for 612 of the remaining viruses. We used the reference sequence for each virus whenever available, or the longest complete genome otherwise. These genomes were used to calculate the relevant genome composition measures described in \cite{mollentze2020identifying}. These were combined with the embeddings to train a series of gradient boosted classification and regression tree models to distinguish between viruses known to infect humans and other viruses. Viruses were randomly split into three datasets, using 70\% for training, 15\% for model calibration, and the remaining 15\% for evaluating model performance \cite{mollentze2020identifying}. This training/calibration/test procedure was repeated 1000 times to assess variability in performance arising from current limited knowledge of the human-infecting virome. We compared models trained on either the original viral genome composition descriptors from \cite{mollentze2020identifying}, or a combination of viral genome composition and embeddings derived from either the observed network or from the imputed network. Finally, the best model by ROC-AUC (the model using viral genome features and embedding-features describing the imputed network) was used to predict the probability of human infection for all 612 viruses. For this purpose, predictions were averaged across the best-performing 10\% of models in which each virus occurred in the test data, a process akin to bagging \cite{babayan2018predicting}. Model performance was re-evaluated while excluding the virus being predicted, to avoid selecting models based on their performance on the virus being predicted.

\clearpage

\bibliographystyle{naturemag}
\bibliography{main}

\begin{thebibliography}{10}

\bibitem{albery2021science}
Albery, G.~F., Becker, D.~J., Brierley, L., Brook, C.~E., Christofferson,
  R.~C., Cohen, L.~E., Dallas, T.~A., Eskew, E.~A., Fagre, A., Farrell, M.~J.,
  et~al.
\newblock {\em Nature microbiology}{ \bf 6}(12), 1483--1492 (2021).

\bibitem{dallas2017predicting}
Dallas, T., Park, A.~W., and Drake, J.~M.
\newblock {\em PLoS computational biology}{ \bf 13}(5), e1005557 (2017).

\bibitem{carlson2019global}
Carlson, C.~J., Zipfel, C.~M., Garnier, R., and Bansal, S.
\newblock {\em Nature ecology \& evolution}{ \bf 3}(7), 1070--1075 (2019).

\bibitem{carlson2020climate}
Carlson, C.~J., Albery, G.~F., Merow, C., Trisos, C.~H., Zipfel, C.~M., Eskew,
  E.~A., Olival, K.~J., Ross, N., and Bansal, S.
\newblock {\em BioRxiv}{ \bf } (2020).

\bibitem{nejati2016denoising}
Nejati, M., Samavi, S., Derksen, H., and Najarian, K.
\newblock {\em Journal of Visual Communication and Image Representation}{ \bf
  36}, 28--39 (2016).

\bibitem{mollentze2020identifying}
Mollentze, N., Babayan, S., and Streicker, D.
\newblock {\em bioRxiv}{ \bf } (2020).

\bibitem{mollentze2020viral}
Mollentze, N. and Streicker, D.~G.
\newblock {\em Proceedings of the National Academy of Sciences}{ \bf 117}(17),
  9423--9430 (2020).

\bibitem{wille2021accurately}
Wille, M., Geoghegan, J.~L., and Holmes, E.~C.
\newblock {\em PLoS Biology}{ \bf 19}(4), e3001135 (2021).

\bibitem{williams2011biology}
Williams, R.~J.
\newblock {\em PLoS One}{ \bf 6}(3), e17645 (2011).

\bibitem{albery2020predicting}
Albery, G.~F., Eskew, E.~A., Ross, N., and Olival, K.~J.
\newblock {\em Nature communications}{ \bf 11}(1), 1--9 (2020).

\bibitem{legendre2013beta}
Legendre, P. and De~C{\'a}ceres, M.
\newblock {\em Ecology letters}{ \bf 16}(8), 951--963 (2013).

\bibitem{poisot2017hosts}
Poisot, T., Gu{\'e}veneux-Julien, C., Fortin, M.-J., Gravel, D., and Legendre,
  P.
\newblock {\em Global ecology and biogeography}{ \bf 26}(8), 942--951 (2017).

\bibitem{olival2017host}
Olival, K.~J., Hosseini, P.~R., Zambrana-Torrelio, C., Ross, N., Bogich, T.~L.,
  and Daszak, P.
\newblock {\em Nature}{ \bf 546}(7660), 646--650 (2017).

\bibitem{grange2021ranking}
Grange, Z.~L., Goldstein, T., Johnson, C.~K., Anthony, S., Gilardi, K., Daszak,
  P., Olival, K.~J., O'Rourke, T., Murray, S., Olson, S.~H., Togami, E., Vidal,
  G., Panel, E., Consortium, P., and Mazet, J. A.~K.
\newblock {\em Proceedings of the National Academy of Sciences}{ \bf 118}(15)
  April  (2021).

\bibitem{babayan2018predicting}
Babayan, S.~A., Orton, R.~J., and Streicker, D.~G.
\newblock {\em Science}{ \bf 362}(6414), 577--580 (2018).

\bibitem{gibb2021data}
Gibb, R., Albery, G.~F., Becker, D.~J., Brierley, L., Connor, R., Dallas,
  T.~A., Eskew, E.~A., Farrell, M.~J., Rasmussen, A.~L., Ryan, S.~J., et~al.
\newblock {\em bioRxiv}{ \bf } (2021).

\bibitem{wardeh2015database}
Wardeh, M., Risley, C., McIntyre, M.~K., Setzkorn, C., and Baylis, M.
\newblock {\em Scientific data}{ \bf 2}(1), 1--11 (2015).

\bibitem{stephens2017global}
Stephens, P.~R., Pappalardo, P., Huang, S., Byers, J.~E., Farrell, M.~J.,
  Gehman, A., Ghai, R.~R., Haas, S.~E., Han, B., Park, A.~W., et~al.
\newblock {\em Ecology}{ \bf 98}(5), 1476 (2017).

\bibitem{shaw2020phylogenetic}
Shaw, L.~P., Wang, A.~D., Dylus, D., Meier, M., Pogacnik, G., Dessimoz, C., and
  Balloux, F.
\newblock {\em Molecular ecology}{ \bf 29}(17), 3361--3379 (2020).

\bibitem{stock2017linear}
Stock, M., Poisot, T., Waegeman, W., and De~Baets, B.
\newblock {\em Scientific reports}{ \bf 7}(1), 1--8 (2017).

\bibitem{strydom2020svd}
Strydom, T., Dalla~Riva, G.~V., and Poisot, T.
\newblock {\em EcoEvoRxiv. November}{ \bf 2} (2020).

\bibitem{Shlens2014TutPri}
Shlens, J.
\newblock {\em arXiv:1404.1100 [cs, stat]}{ \bf } (2014).

\bibitem{bezanson2012julia}
Bezanson, J., Karpinski, S., Shah, V.~B., and Edelman, A.
\newblock {\em arXiv preprint arXiv:1209.5145}{ \bf } (2012).

\bibitem{poisot2019ecologicalnetworks}
Poisot, T., B{\'e}lisle, Z., Hoebeke, L., Stock, M., and Szefer, P.
\newblock {\em Ecography}{ \bf 42}(11), 1850--1861 (2019).

\bibitem{upham2019inferring}
Upham, N.~S., Esselstyn, J.~A., and Jetz, W.
\newblock {\em PLoS biology}{ \bf 17}(12), e3000494 (2019).

\bibitem{winter2017rentrez}
Winter, D.~J.
\newblock {\em The R Journal}{ \bf 9}(2), 520--526 (2017).

\bibitem{teitelbaum2020estimators}
Teitelbaum, C.~S., Amoroso, C.~R., Huang, S., Davies, T.~J., Rushmore, J.,
  Drake, J.~M., Stephens, P.~R., Byers, J.~E., Majewska, A.~A., and Nunn, C.~L.
\newblock {\em Ecography}{ \bf 43}(9), 1316--1328 (2020).

\bibitem{nunn2003comparative}
Nunn, C.~L., Altizer, S., Jones, K.~E., and Sechrest, W.
\newblock {\em The American Naturalist}{ \bf 162}(5), 597--614 (2003).

\bibitem{ezenwa2006host}
Ezenwa, V.~O., Price, S.~A., Altizer, S., Vitone, N.~D., and Cook, K.~C.
\newblock {\em Oikos}{ \bf 115}(3), 526--536 (2006).

\bibitem{huang2015parasite}
Huang, S., Drake, J.~M., Gittleman, J.~L., and Altizer, S.
\newblock {\em Evolution}{ \bf 69}(3), 621--630 (2015).

\bibitem{luis2013comparison}
Luis, A.~D., Hayman, D.~T., O'Shea, T.~J., Cryan, P.~M., Gilbert, A.~T.,
  Pulliam, J.~R., Mills, J.~N., Timonin, M.~E., Willis, C.~K., Cunningham,
  A.~A., et~al.
\newblock {\em Proceedings of the Royal Society B: Biological Sciences}{ \bf
  280}(1756), 20122753 (2013).

\bibitem{freckleton2002phylogenetic}
Freckleton, R.~P., Harvey, P.~H., and Pagel, M.
\newblock {\em The American Naturalist}{ \bf 160}(6), 712--726 (2002).

\bibitem{orme2012caper}
Orme, D., Freckleton, R., Thomas, G., Petzoldt, T., Fritz, S., Isaac, N.,
  Pearse, W., et~al.
\newblock {\em R package version 0.5}{ \bf 2}, 458 (2012).

\bibitem{pagel1999inferring}
Pagel, M.
\newblock {\em Nature}{ \bf 401}(6756), 877--884 (1999).

\bibitem{washburne2019phylofactorization}
Washburne, A.~D., Silverman, J.~D., Morton, J.~T., Becker, D.~J., Crowley, D.,
  Mukherjee, S., David, L.~A., and Plowright, R.~K.
\newblock {\em Ecological Monographs}{ \bf 89}(2), e01353 (2019).

\bibitem{dalla2016exploring}
Dalla~Riva, G.~V. and Stouffer, D.~B.
\newblock {\em Oikos}{ \bf 125}(4), 446--456 (2016).

\bibitem{woolhouse_epidemiological_2018}
Woolhouse, M. E.~J. and Brierley, L.
\newblock {\em Scientific Data}{ \bf 5}, 180017 February  (2018).

\bibitem{ogunkoya_serological_1990}
Ogunkoya, A.~B., Beran, G.~W., Umoh, J.~U., Gomwalk, N.~E., and Abdulkadir,
  I.~A.
\newblock {\em Transactions of the Royal Society of Tropical Medicine and
  Hygiene}{ \bf 84}(6), 842--845 December  (1990).

\bibitem{mendez2012genome}
Mendez-Rios, J.~D., Martens, C.~A., Bruno, D.~P., Porcella, S.~F., Zheng,
  Z.-M., and Moss, B.
\newblock {\em PLoS One}{ \bf 7}(4), e34604 (2012).

\bibitem{kuroya_newborn_1953}
Kuroya, M. and Ishida, N.
\newblock {\em Yokohama Medical Bulletin}{ \bf 4}(4), 217--233 August  (1953).

\bibitem{maruyama_survey_1983}
Maruyama, S.
\newblock {\em Journal of the Japan Veterinary Medical Association}{ \bf
  36}(6), 330--333 (1983).

\bibitem{centers_for_disease_control_cdc_update_1990}
{Centers for Disease Control (CDC)}.
\newblock {\em MMWR. Morbidity and mortality weekly report}{ \bf 39}(13), 221
  April  (1990).

\bibitem{olaleye1990survey}
Olaleye, O., Omilabu, S., Ilomechina, E., and Fagbami, A.
\newblock {\em Comparative immunology, microbiology and infectious diseases}{
  \bf 13}(1), 35--39 (1990).

\bibitem{lvov_isolation_1980}
L'vov, D.~K., Kostiukova, M.~A., Pak, T.~P., and Gromashevskiĭ, V.~L.
\newblock {\em Voprosy Virusologii}{ \bf }(1), 61--62 February  (1980).

\end{thebibliography}

\clearpage
\newpage

\section*{Figures and Tables}

\clearpage

\renewcommand{\figurename}{Extended Data Figure}
\setcounter{figure}{0} 

\clearpage

\begin{figure}
    \centering
    \includegraphics[width = 15cm]{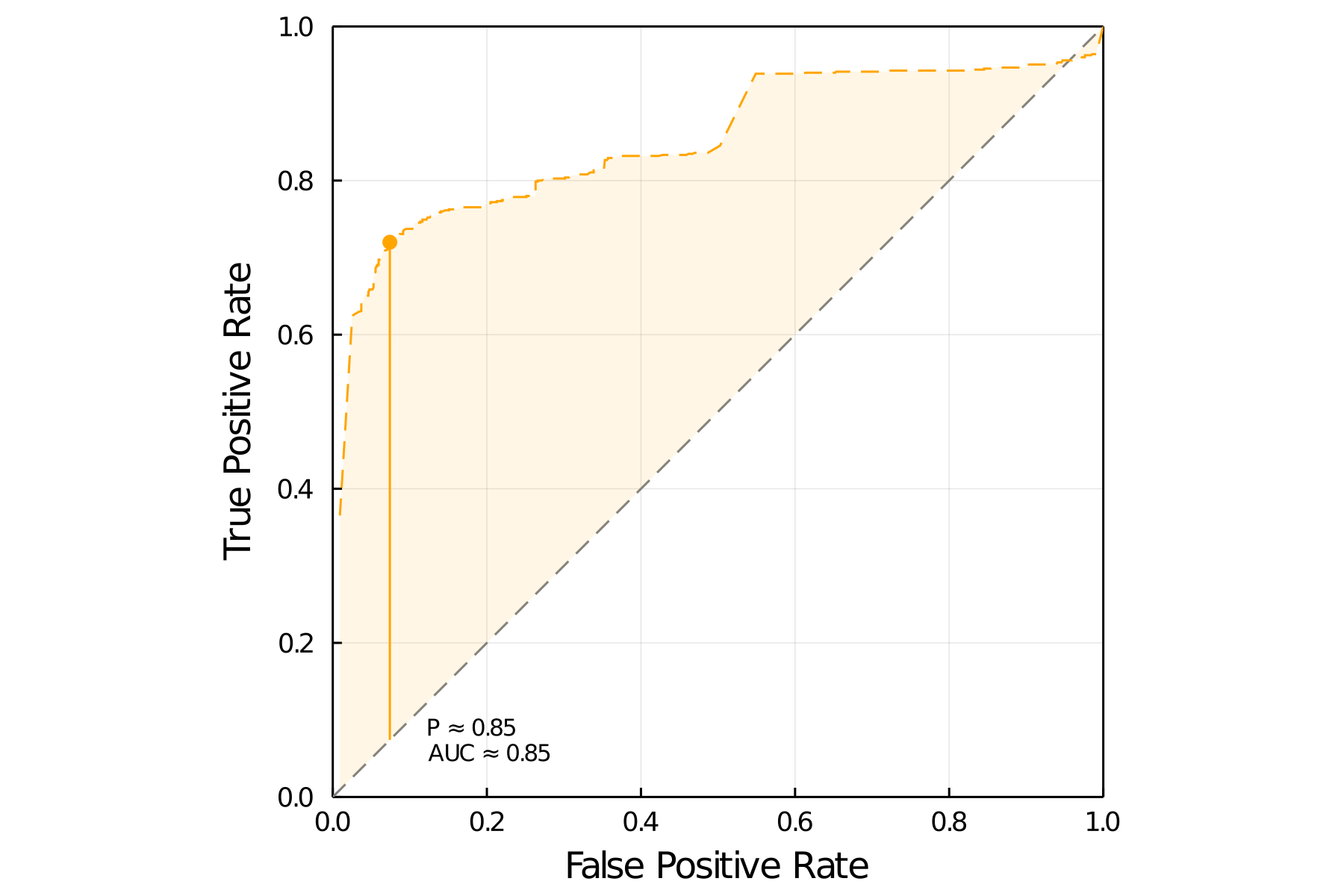}
    \caption{\textbf{Receiver operating characteristic (ROC) curve for the best model}. All models (from SVD rank 1 to 20, and using three linear filtering parameterization) have been compared on the same training/validation dataset by measuring the area under the ROC curve. The best model has an area under the ROC curve (AUC) of 0.85. The threshold turning the continuous prediction of LF-SVD into a binary classification was picked to simultaneously maximize the true positive rate and minimize the false positive rate, in practice by maximizing Youden's informedness. By coincidence, this threshold is also approximately equal to 0.85 in the best model.}
    \label{fig:bestrocauc}
\end{figure}

\clearpage

\begin{figure}
    \centering
    \includegraphics[width = 15cm]{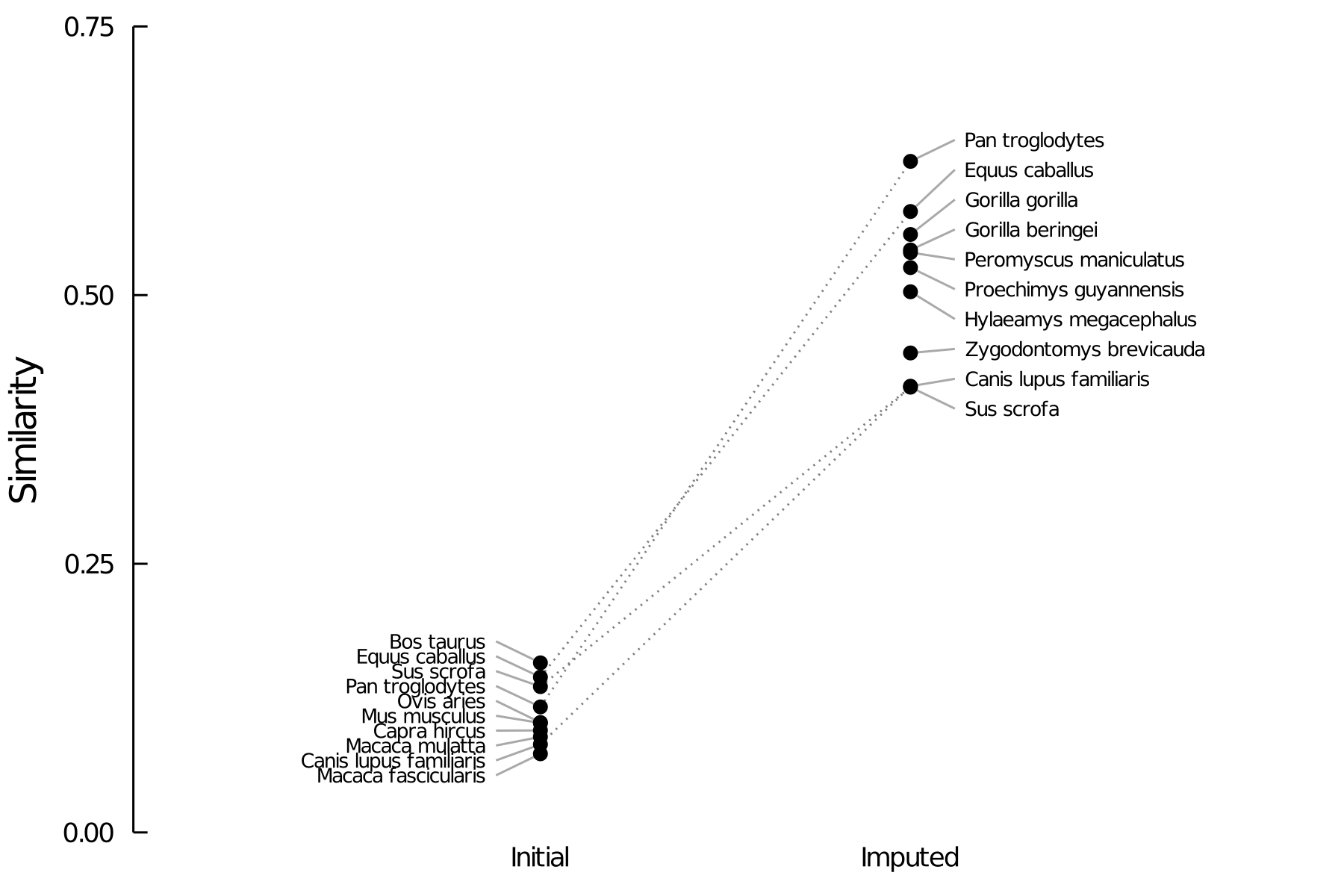}
    \caption{\textbf{Pairwise host-similarity to \textit{Homo sapiens} changes post imputation}. The ten hosts with the most viral overlap to \textit{Homo sapiens} (as measured by Jaccard similarity) tend to be livestock. By contrast, the ten most similar hosts after imputation are mostly composed of primates and rodents, which suggests that LF-SVD is able to overcome taxonomic bias in the original dataset.}
    \label{fig:human10share}
\end{figure}

\clearpage

\begin{figure}
    \centering
    \includegraphics[width = 13cm]{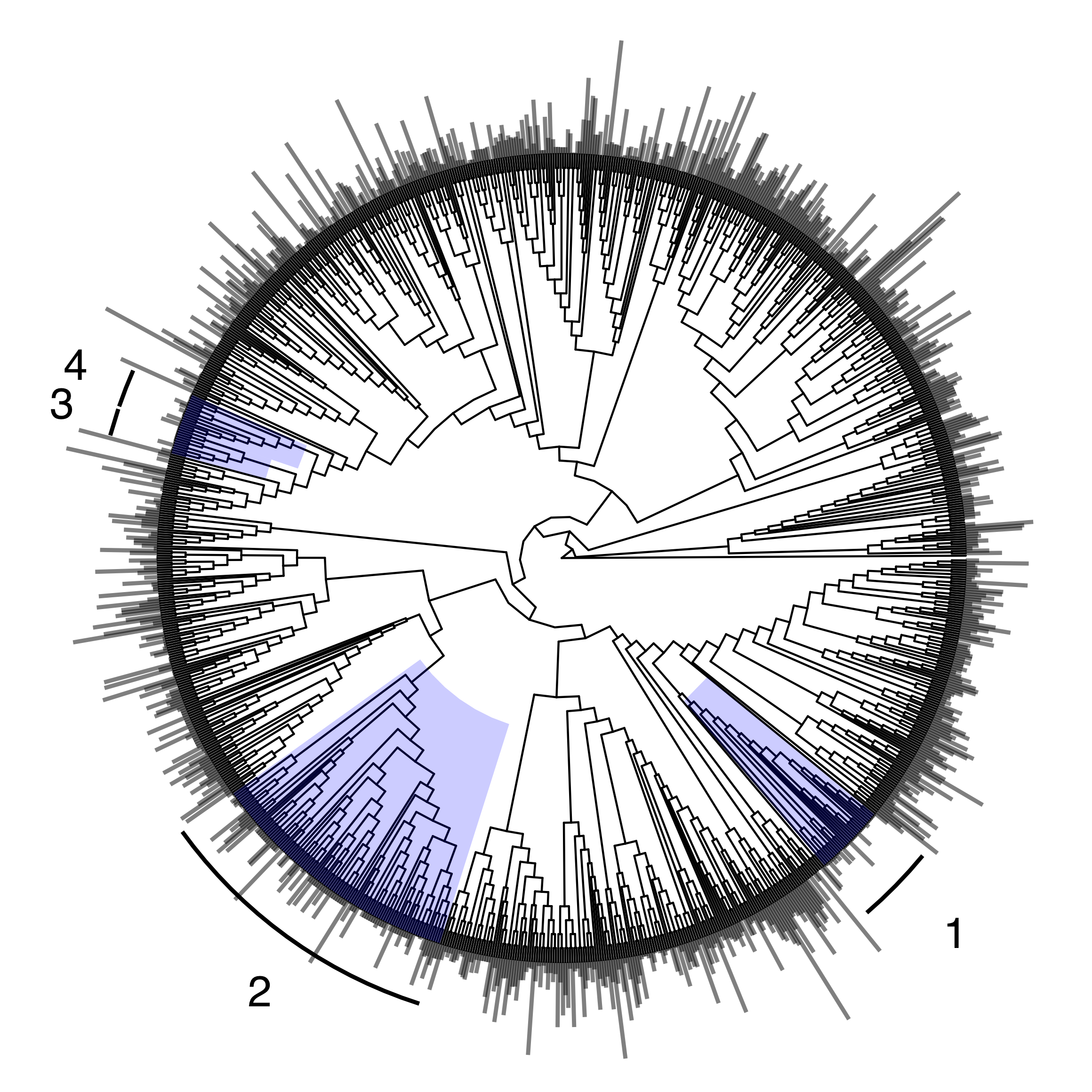}
    \caption{\textbf{Phylogenetic bias in missing viruses.} Phylogenetic factorization determined that the majority of species have no phylogenetic signal in the number of missing viruses estimated by the LF-SVD model, with the exception of a handful of small clades that included cetaceans (clade 1), a mostly insectivorous subclade of the Yangochiroptera (clade 2), and two small rodent clades (clades 3 and 4), all of which have significantly fewer than average. }
    \label{fig:phylofactor}
\end{figure}

\clearpage

\begin{figure}
    \centering
    \includegraphics[width = 16cm]{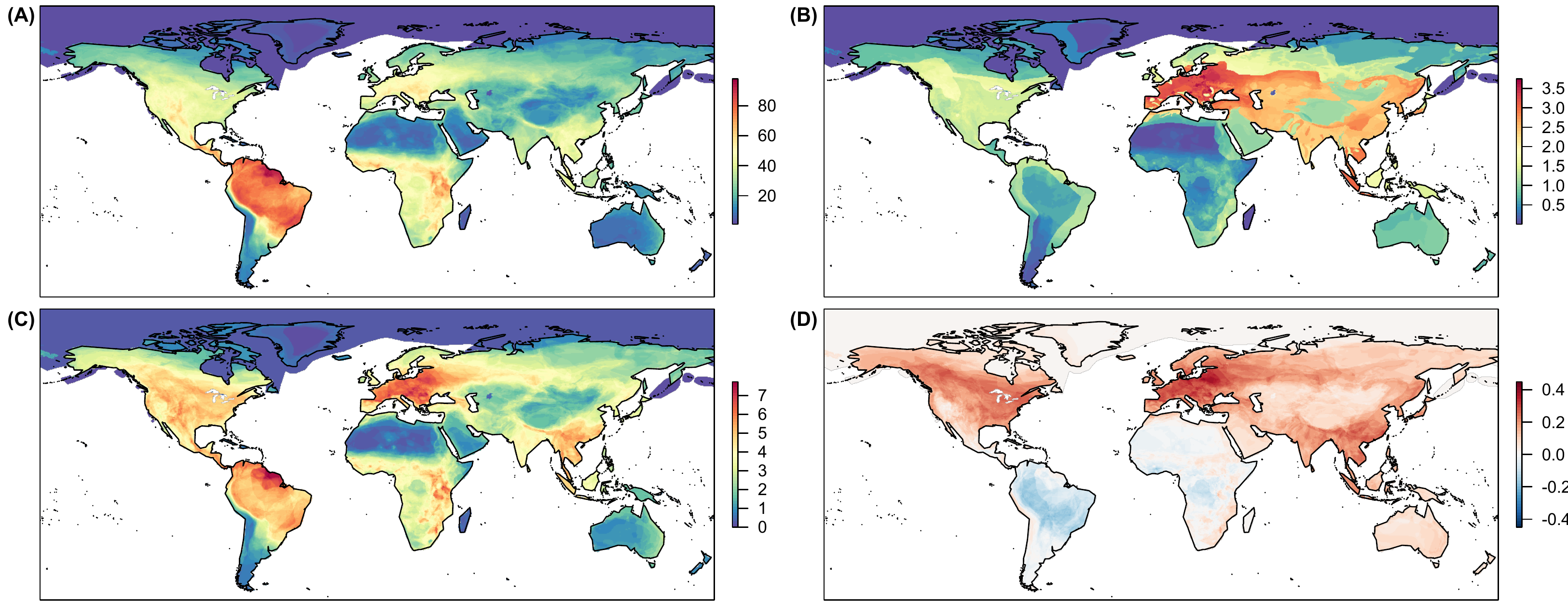}
    \caption{\textbf{Geographic biases in sparsity are reduced, but not entirely eliminated, by the imputation model.} Host diversity (number of species) in the CLOVER dataset closely tracks true patterns of global biodiversity (A), but the total number of interactions recorded does not (B), due to a high degree of sampling bias. Model-based predictions of undiscovered interactions (C) much more closely track true biodiversity gradients, but likely underestimates in South America and Africa due to sparsity (D). Numbers of interactions are given in thousands (B,C). Hotspots in (D) are given as the difference between the number of undiscovered interactions and underlying host diversity, both rescaled between 0 and 1.}
    \label{fig:geogbias}
\end{figure}

\clearpage

\begin{figure}
    \centering
    \includegraphics[width = \textwidth]{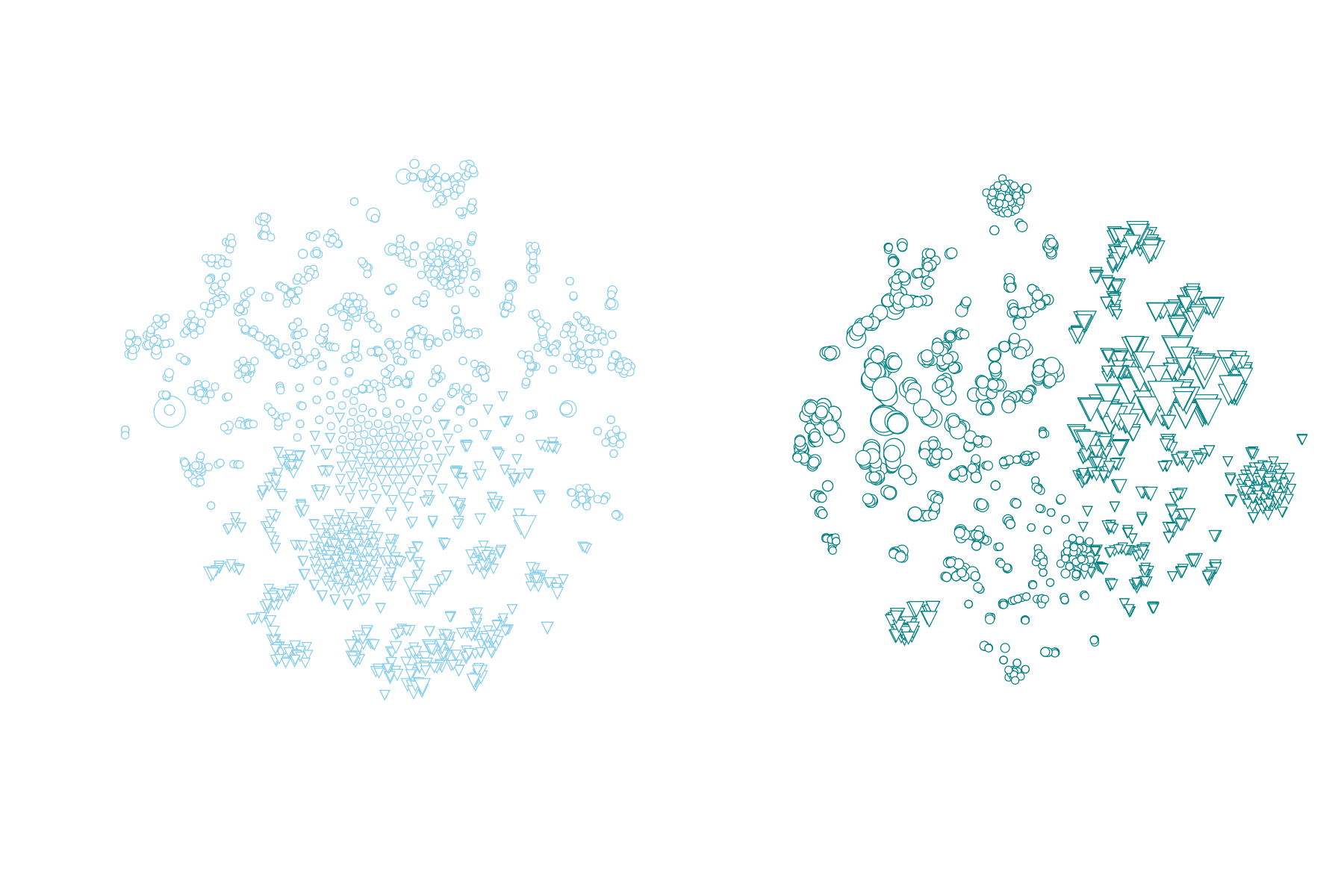}
    \caption{\textbf{The global virome, pre- and post-imputation.} Network layouts reflect the first two dimensions of a tSNE embedding on four dimensions, wherein the positions of nodes where initially picked based on a principal components analysis. Hosts are shown as circles and viruses as downwards-pointing triangles, and the relative size of each point scales linearly with degree (using the same scale for both figures, \textit{i.e.} two nodes with the same degree will have the same size in the left and right panels).}
    \label{fig:dotty}
\end{figure}

\clearpage

\begin{figure}
    \centering
    \includegraphics[width = 13cm]{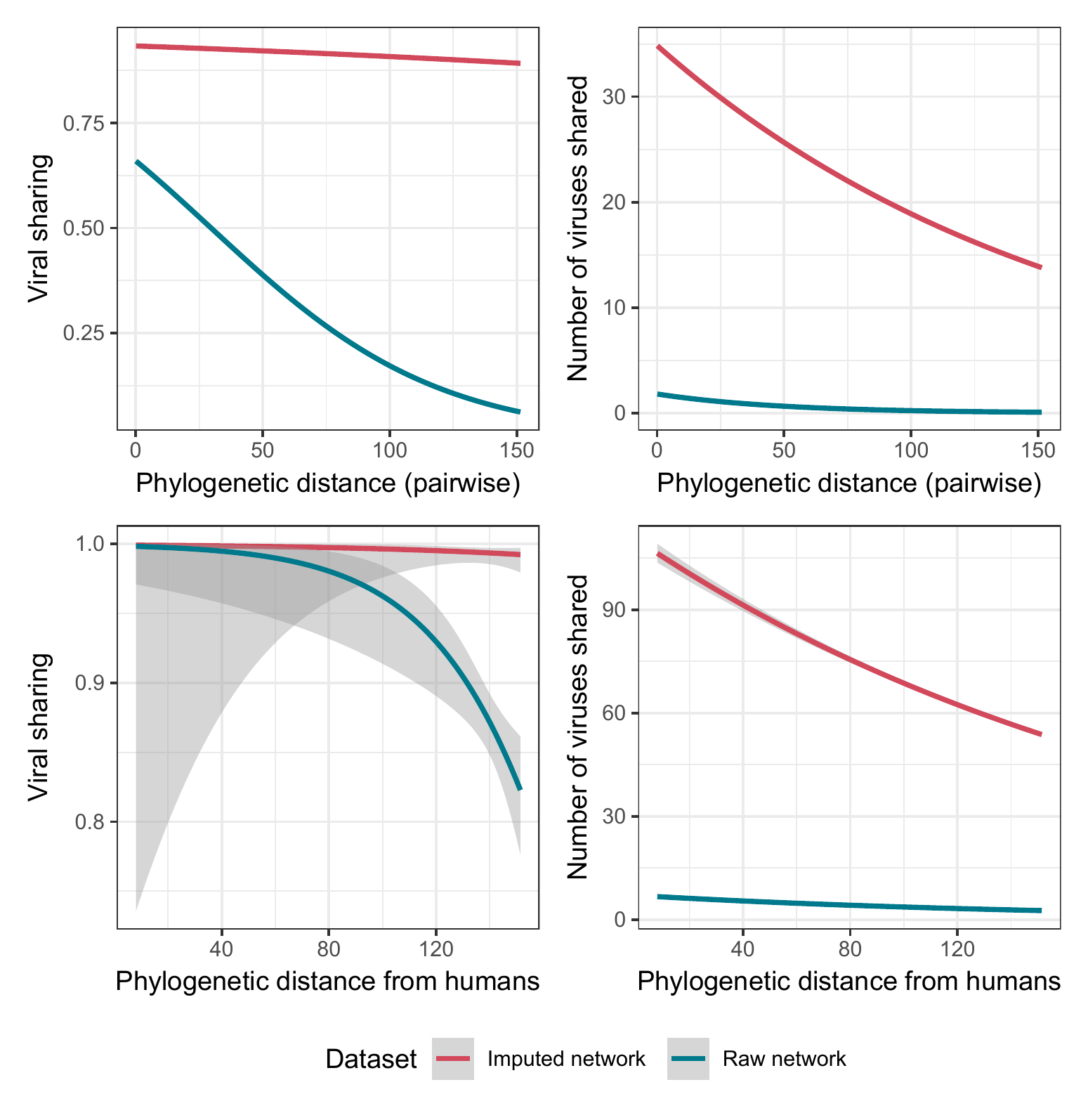}
    \caption{\textbf{Evolutionary signals dominate viral sharing.} Phylogenetic distance among all mammals (top) or from humans (bottom) structure viral sharing measured as a binary trait (left) or based on the total number of shared viruses (right). In the imputed network, mosts hosts have many more ; as a result, phylogenetic distance is less informative of whether hosts share viruses, because most hosts share at least one virus, but the phylogenetic signal of the count data is much stronger. Curves are given as generalized linear model smooths, with a Poisson distribution for count data and a binomial distribution with a logit link function for viral sharing.}
    \label{fig:phylohuman}
\end{figure}

\clearpage

\begin{figure}
    \centering
    \includegraphics[width = 15cm]{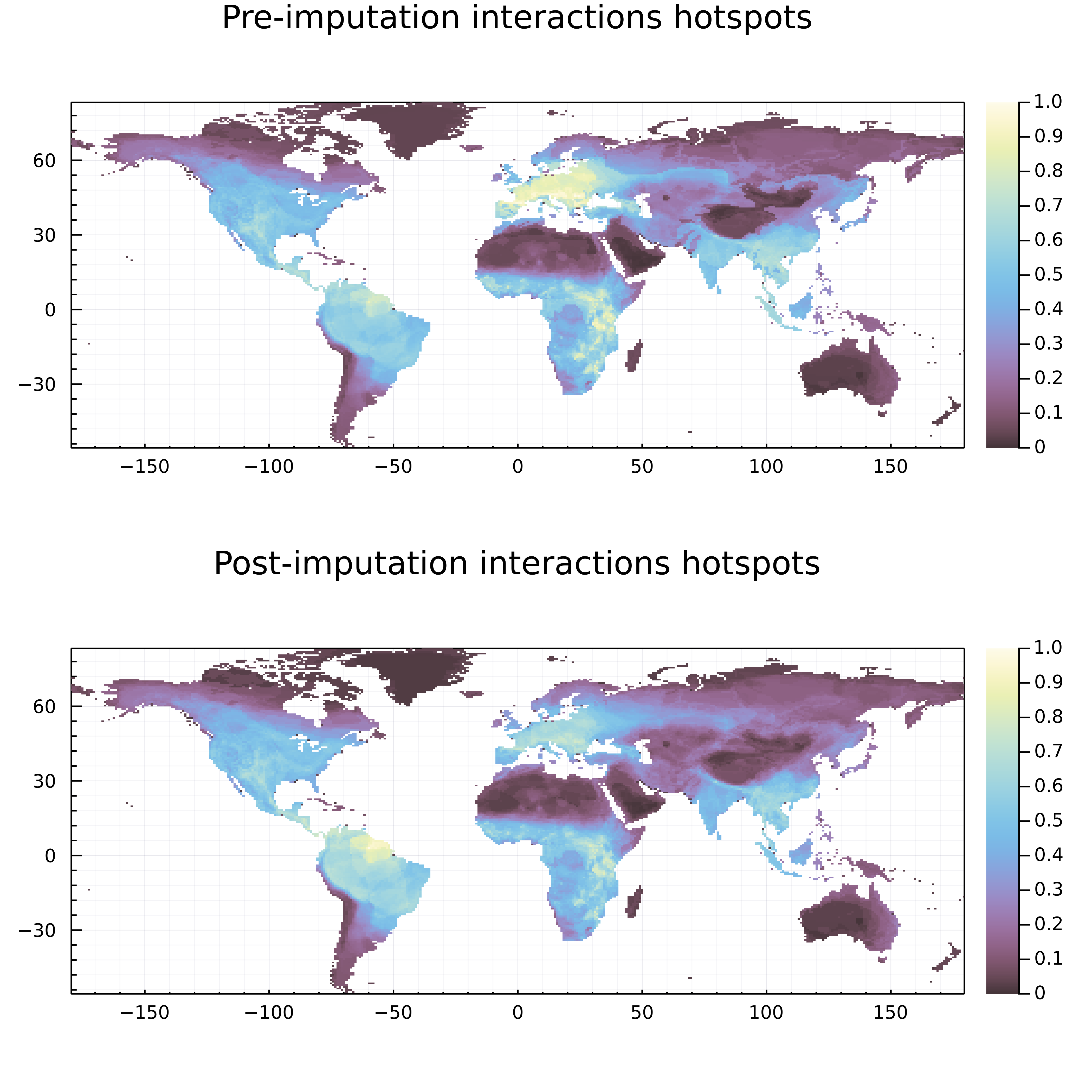}
    \caption{\textbf{Imputation changes the spatial location of viral uniqueness hostpots.} The top and bottom panel show, respectively, the LCBD calculated on viral community composition before and after imputation. Although sampling biases in the original dataset (notably an over-sampling of livestock viruses) puts a lot of emphasis on Europe, the main hostpot post-imputation is in the Amazon. Linear regression between the two layers reveals that the bias reduction only has a moderate effect on the overall relative patterns (constrained to have a 0 intercept; $t = 1080$; $R^2$ = $0.92$.}
    \label{fig:lcbd-prepost}
\end{figure}

\clearpage

\begin{figure}
    \centering
    \includegraphics[width = 15.5cm]{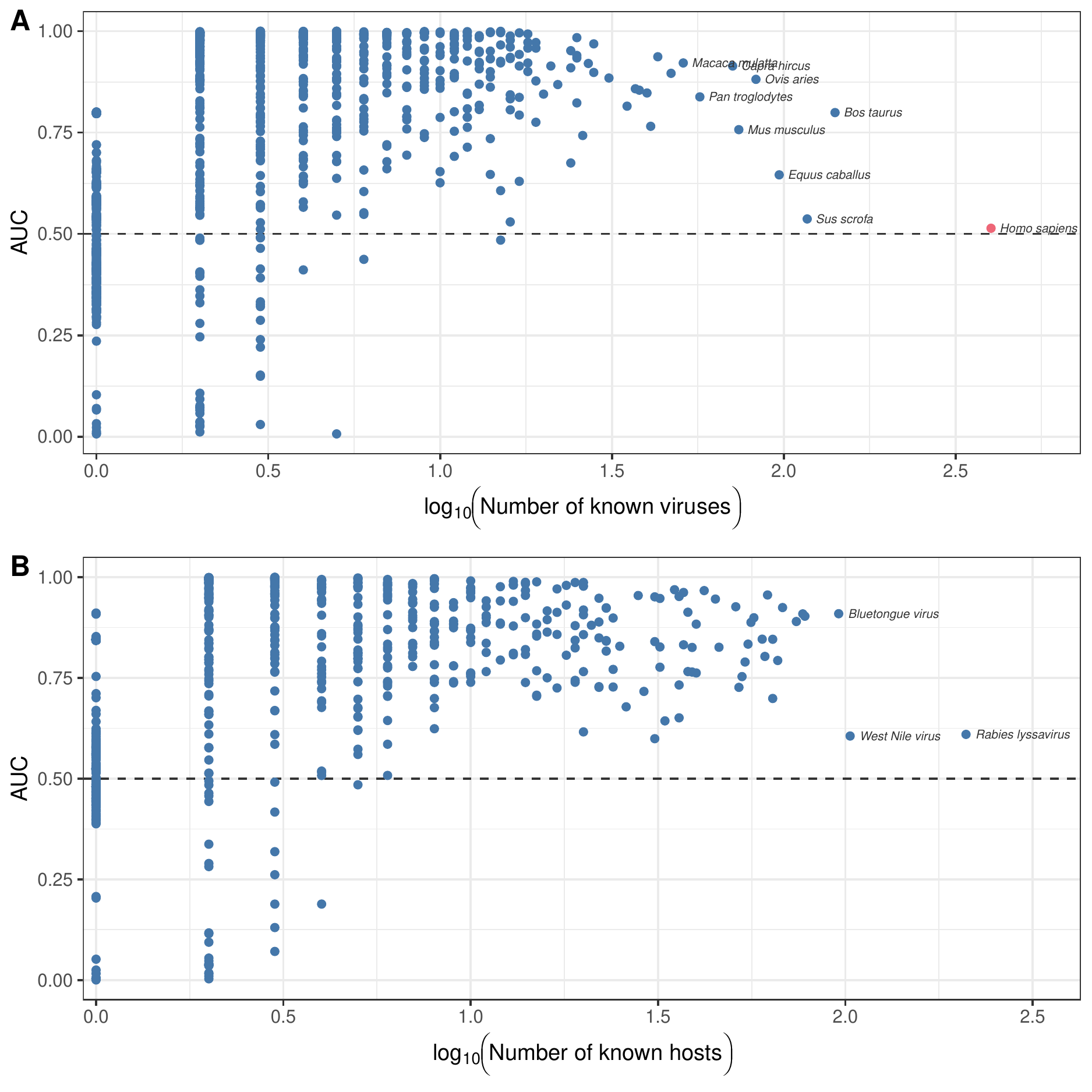}
    \caption{\textbf{Predictive performance of LF-SVD generally increases with increased connectivity.} Points represent individual host species, and show the probability that a randomly sampled virus known to infect that host will be ranked above a randomly sampled virus which has not been observed to do so (measured as the area under the receiver operating characteristic curve [AUC]). While hosts subject to extreme study bias such as humans cannot be predicted, this does not appear to degrade performance on other species.}
    \label{fig:connectivity}
\end{figure}

\clearpage


\clearpage

\begin{figure}
    \centering
    \includegraphics[width = 15.5cm]{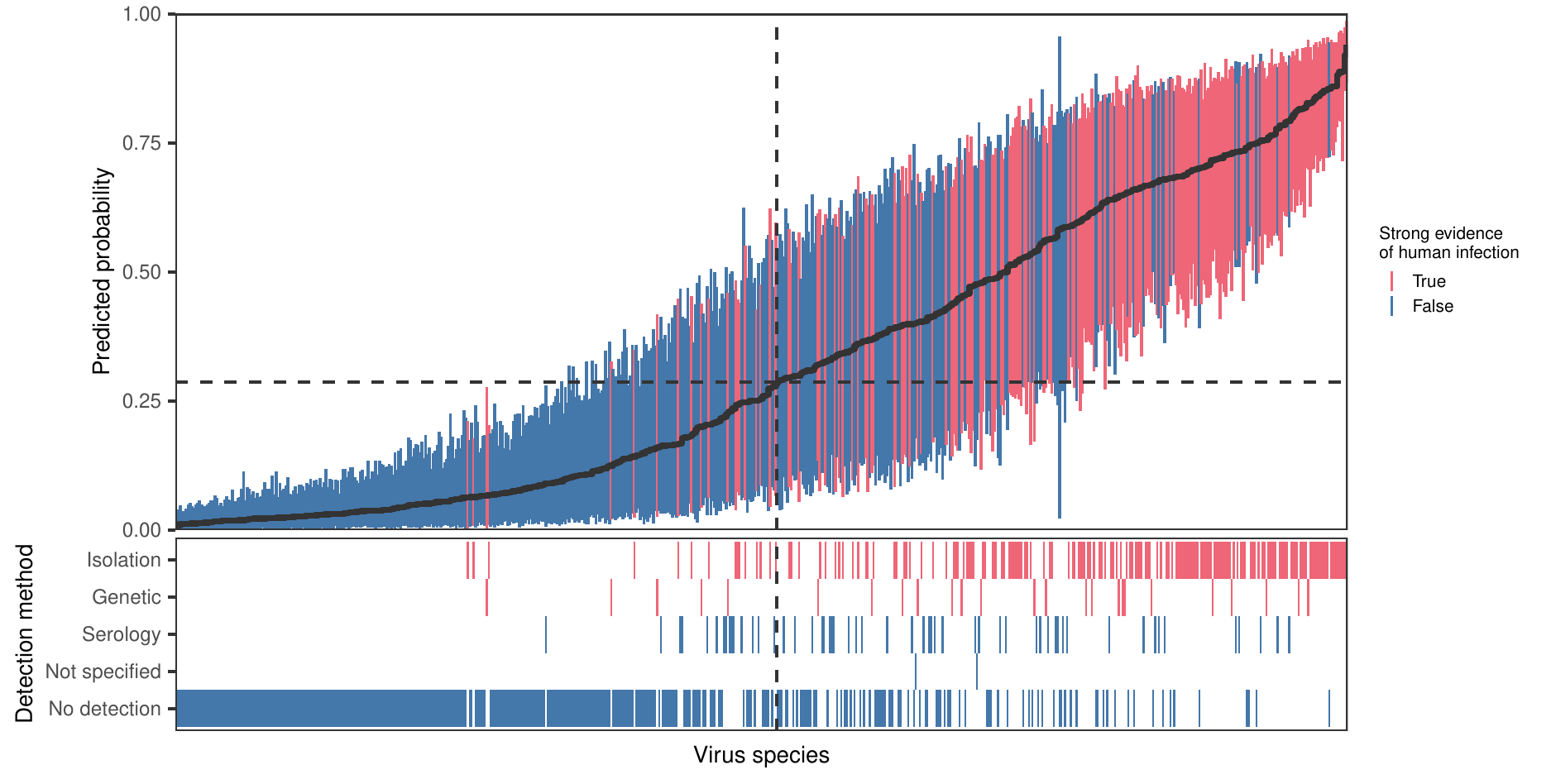}
    \caption{\textbf{Ranking viruses by their predicted probability of human infection accurately predicts known infections.} Viruses are arranged by the mean prediction produced by a bagged version of the model trained on both genome composition features and an embedding representing the imputed network (black line). Error bars show the region containing 95\% of the predictions used for bagging. Dashed lines highlight the cut-off which maximizes informedness (Youden's J) when converting mean predicted probabilities to binary predictions. A second panel shows the most reliable detection method providing evidence of human infection for each virus in the CLOVER database. For the purposes of model training, viruses linked to humans through serological detections only or where the detection method was unspecified were labelled as negative; the model nevertheless identifies the majority of these as human-infecting.}
    \label{fig:serology}
\end{figure}

\clearpage

\renewcommand{\tablename}{Extended Data Table}
\setcounter{table}{0} 

\begin{table}[ht]
\caption{\textbf{Model performance for the top 10 models by AUC.} Metrics include the AUC and cutoff (expressed as a pseudo-probability), the true positive and true negative rates (TPR, TNR), the positive and negative predictive values (PPV, NPV), the false negative and positive rates (FNR, FPR), the false discovery and false omission rates (FDR, FOR), the critical success index (CSI), accuracy (ACC), and Youden's J. \\}
\begin{tabular}{|l|l|llllll|}
\hline
model       & rank & AUC   & cutoff & TPR   & TNR   & PPV   & NPV   \\ \hline
connectance & 12   & 0.849 & 0.846  & 0.72  & 0.925 & 0.906 & 0.769 \\
connectance & 11   & 0.846 & 0.908  & 0.684 & 0.936 & 0.914 & 0.75  \\
connectance & 17   & 0.844 & 0.929  & 0.692 & 0.935 & 0.913 & 0.754 \\
connectance & 8    & 0.842 & 0.705  & 0.701 & 0.895 & 0.868 & 0.751 \\
hybrid      & 12   & 0.841 & 0.707  & 0.703 & 0.877 & 0.851 & 0.748 \\
connectance & 14   & 0.839 & 0.902  & 0.7   & 0.929 & 0.907 & 0.758 \\
hybrid      & 11   & 0.837 & 0.82   & 0.647 & 0.918 & 0.888 & 0.723 \\
connectance & 5    & 0.836 & 0.931  & 0.66  & 0.94  & 0.916 & 0.735 \\
connectance & 7    & 0.836 & 0.948  & 0.655 & 0.957 & 0.939 & 0.735 \\
connectance & 16   & 0.835 & 0.961  & 0.667 & 0.945 & 0.923 & 0.741 \\ \hline
\end{tabular}

\quad \\
(Continued:)
\\

\begin{tabular}{|l|l|lllllll|}
\hline
model       & rank & FNR   & FPR   & FDR   & FOR   & CSI   & ACC   & J     \\ \hline
connectance & 12   & 0.28  & 0.074 & 0.093 & 0.23  & 0.669 & 0.823 & 0.645 \\
connectance & 11   & 0.315 & 0.063 & 0.085 & 0.25  & 0.643 & 0.811 & 0.621 \\
connectance & 17   & 0.307 & 0.064 & 0.086 & 0.245 & 0.649 & 0.814 & 0.627 \\
connectance & 8    & 0.298 & 0.104 & 0.131 & 0.248 & 0.634 & 0.798 & 0.596 \\
hybrid      & 12   & 0.296 & 0.122 & 0.148 & 0.251 & 0.626 & 0.79  & 0.581 \\
connectance & 14   & 0.299 & 0.07  & 0.092 & 0.241 & 0.653 & 0.815 & 0.629 \\
hybrid      & 11   & 0.352 & 0.081 & 0.111 & 0.276 & 0.598 & 0.783 & 0.566 \\
connectance & 5    & 0.339 & 0.059 & 0.083 & 0.264 & 0.623 & 0.8   & 0.6   \\
connectance & 7    & 0.344 & 0.042 & 0.06  & 0.264 & 0.628 & 0.806 & 0.613 \\
connectance & 16   & 0.332 & 0.054 & 0.076 & 0.258 & 0.632 & 0.807 & 0.613 \\ \hline
\end{tabular}    
\label{fig:svd-auc}
\end{table}

\clearpage

\begin{table}[!ht]
    \caption{\textbf{Imputation reduces the effect of sampling bias.} To explore whether network imputation via LF-SVD is extrapolating existing research biases, we conducted a set of comparative analyses investigating the how the explanatory power of sampling effort on viral species richness changes after network imputation. We find that after imputation, the slope of the relationship ($\beta$) decreases, and sampling effort explains less of the variance in viral richness ($R^2$), suggesting that imputation via LF-SVD is not merely recapitulating the observed sampling effort per host. Statistics are given for a phylognetic generalized linear model fit with the maximum likelihood estimate of Pagel's $\lambda$. Predictors and responses were log-10 transformed prior to analyses.} 
    \centering
    \quad \\
    \begin{tabular}{|c|c|c|c|c|c|c|} \hline
        Viral richness & Publications & $\beta$ & S.E. & $R^2$ & $\lambda$ & $\lambda$ 95\% CI \\ \hline
        Observed & Cites (all) & 0.53 & 0.02 & 0.46 & 0.59 & (0.47-0.69) \\ 
        LF-SVD imputed & Cites (all) & 0.39 & 0.02 & 0.23 & 0.59 & (0.45-0.72) \\
        \hline
        Observed & Cites (virus-related) & 0.71 & 0.02 & 0.54 & 0.45 & (0.31-0.58) \\ 
        LF-SVD imputed & Cites (virus-related) & 0.47 & 0.03 & 0.22 & 0.60 & (0.46-0.71) \\ \hline
    \end{tabular}
    \label{tab:citations}
\end{table}

\begin{table}[!ht]
\caption{\textbf{Phylofactorization of missing viruses.} Significant clades identified from a phylogenetic factorization of missing virus counts. Included taxa are listed alongside the number of species and the mean number of missing viruses for each clade in comparison to the paraphyletic remainder. Clade codes match ED Figure \ref{fig:phylofactor}. \\ }
\centering
\begin{tabular}{|c|>{\centering\arraybackslash}m{9cm}|c|c|c|}
\hline
Clade & Included taxa & \textit{n} & clade & other \\ \hline
1 & Ziphiidae, Physeteridae, Phocoenidae, Monodontidae, Delphinidae, Eschrichtiidae, Balaenopteridae, Balaenidae & 30 & 18 & 71 \\ \hline
2 & Nycteridae, Emballonuridae, Natalidae, Molossidae, Vespertilionidae	& 109 & 43 & 73 \\ \hline
3 & \textit{Calomys}, \textit{Graomys}, \textit{Phyllotis}, \textit{Loxodontomys}, \textit{Abrothrix} & 11 & 11 & 70 \\ \hline
4 & \textit{Bibimys}, \textit{Oxymycterus}, \textit{Necromys}, \textit{Akodon}, \textit{Thaptomys} & 15 & 16 & 70 \\ \hline

\hline
\end{tabular}
\label{tab:phylofactor}
\end{table}

\begin{table}[!ht]
    \caption{\textbf{Phylogenetic signal in viral sharing, pre-and post-imputation.} Statistics are given for a generalized linear model fit with a binomial distribution for the outcome variable (whether any viruses at all are shared between two hosts). Significance given as *** indicates p < 0.001.}
    \centering 
    \quad \\
    \begin{tabular}{|c|c|c|c|c|c|}
    \hline
      Sharing & Data source & $\beta$ & S.E. & p & $R^2$ (adj.) \\       \hline
      Pairwise (all hosts) & Pre-imputation & 2.23 e-2 & 8.44 e-05 & *** & 9.8\% \\
      Pairwise (all hosts) & Post-imputation & 3.50 e-03 & 1.30 e-4 & *** & 0.07\% \\ \hline
      With \textit{Homo sapiens} & Pre-imputation & 3.32 e-2 & 1.07 e-2 & *** & 2.4\% \\
      With \textit{Homo sapiens} & Post-imputation & 1.46 e-2 & 2.30 e-2 & 0.524 & -0.09\% \\ \hline
    \end{tabular}
    \label{tab:phylohuman1}
\end{table}

\begin{table}[!ht]
    \caption{\textbf{Phylogenetic signal in number of viruses shared, pre-and post-imputation.} Statistics are given for a generalized linear model fit with a Poisson distribution for the outcome variable. Significance given as *** indicates p < 0.001.}
    \centering 
    \quad \\
    \begin{tabular}{|c|c|c|c|c|c|}
    \hline
      Sharing & Data source & $\beta$ & S.E. & p & $R^2$ (adj.) \\       \hline
        Pairwise (all hosts) & Pre-imputation & -2.04 e-02 & 4.97 e-05 & *** & 8.5\%\\
        Pairwise (all hosts) & Post-imputation & -6.12 e-03 & 7.20 e-06 & *** & 3.0\%\\ \hline
        With \textit{Homo sapiens} & Pre-imputation & 6.53 e-03 & 4.50 e-04 & *** & 2.2\% \\
        With \textit{Homo sapiens} & Post-imputation & 4.76 e-0.3 & 1.08 e-04 & *** & 3.8\% \\ \hline
    \end{tabular}
    \label{tab:phylohuman2}
\end{table}

\begin{table}[!ht]
    \caption{\textbf{The top 10 predicted (novel) zoonotic links in the post-imputation network.} Evidence of interaction generated by the imputation model is contrasted against prior predictions by \cite{mollentze2020identifying}, who implemented a model that successfully predicts zoonotic potential from viral genome composition bias.}
    \centering
    \quad \\
    \begin{tabular}{|c|c|c|c|}
    \hline 
         Virus & Family & Evidence & Prior risk assignment \\
         \hline
\textit{Canine mastadenovirus A} & \textit{Adenoviridae} & 275.6808 & Medium \\
\textit{Simian mastadenovirus A} & \textit{Adenoviridae} &	242.8597 & --\\
\textit{Panine gammaherpesvirus 1} & \textit{Herpesviridae} &	201.9715 & -- \\
\textit{Phocid alphaherpesvirus 1} & \textit{Herpesviridae} & 191.4652 & High \\
\textit{Carnivore protoparvovirus 1} & \textit{Parvoviridae} & 191.2557 & High \\
\textit{Torque teno virus 14} & \textit{Annelloviridae} & 187.3940 & High \\
\textit{Torque teno virus 4} & \textit{Annelloviridae} & 187.3940 & Medium \\
\textit{Panine betaherpesvirus 2} & \textit{Herpesviridae} & 187.3940 & High \\
\textit{Torque teno virus 23} & \textit{Annelloviridae} & 187.3940 & High \\
\textit{Torque teno virus 2} & \textit{Annelloviridae} & 182.4210 & Medium \\
\hline
    \end{tabular}
    \label{tab:svd-zoon}
\end{table}

\begin{table}[!ht]
    \caption{\textbf{The top 20 predicted (novel) zoonotic viruses in the extended model.} All are classified as ``very high'' risk by the combined model, which uses both viral genome compositions and imputed network embeddings. Prior risk assignments from \cite{mollentze2020identifying} are also given where possible. (*: Indicates that a virus has serological evidence of human infection in CLOVER, which was not included as a positive in the genomic model, but was considered evidence of association in the mammal-virus network; however, note that \textit{Homo sapiens} and its associations were dropped before generating embeddings. $\dagger$: Indicates that a virus has recorded evidence of human infection in CDC's ArboCat, though original source literature is not traceable. $\ddag$: Indicates that a virus is accepted as a human virus by \cite{woolhouse_epidemiological_2018}. Caption continues next page.)}
    \centering
    \quad \\
    {\renewcommand{\arraystretch}{1.5}
    \begin{tabular}{|l|m{0.12\linewidth}|>{\raggedright\arraybackslash}m{0.23\linewidth}|c|m{0.09\linewidth}|}
         \hline
        \textbf{Virus} & \textbf{Virus family (-viridae)} & \textbf{Animal hosts (number of spp.)} & \textbf{Prob.} & \textbf{Prior risk} \\
         \hline
        \textit{Lagos bat lyssavirus}\textsuperscript{[a]} & Rhabdo & Chiroptera (10), Carnivora (3), Rodentia (1) & 0.856 & Very high \\
        \textit{Tacaribe mammarenavirus}* & Arena & Chiroptera (9), Rodentia (1) & 0.793 & High \\
        \textit{Rio Bravo virus}* & Flavi & Chiroptera (19) & 0.779 & Medium \\
        \textit{Dera Ghazi Khan orthonairovirus}*	& Nairo & Rodentia (4), Artiodactyla (2) & 0.755 & Medium \\
        \textit{Wad Medani virus} & Reo & Artiodactyla (6), Rodentia (4) & 0.750 & Medium \\
        \textit{Enterovirus E}$\ddag$ & Picorna & Artiodactyla (1), Primates (1) & 0.745 & Low \\
        \textit{Phocine morbillivirus} & Paramyxo & Carnivora (22) & 0.741 & High \\
        \textit{Bimiti orthobunyavirus}*$\dagger$ & Peribunya & Chiroptera (5), Rodentia (4), Perissodactyla (1) & 0.734 & High \\
        \textit{Bujaru phlebovirus}*$\dagger$ & Phenui & \textit{Proechimys guyannensis}	(Rodentia) & 0.733 & Very high \\
        \textit{Ectromelia virus}\textsuperscript{[b]} & Pox & Rodentia (3), Carnivora (1) & 0.701 & High \\ 
         \hline
    \end{tabular}}
    \label{tab:nardus-zoon}
\end{table}

\clearpage

\addtocounter{table}{-1}
\renewcommand{\thetable}{\arabic{table} (continued)}

\begin{table}[!ht]
    \caption{Caption cont'd. \textsuperscript{[a]} Serological evidence recorded from four human samples \cite{ogunkoya_serological_1990}.
    \textsuperscript{[b]} A strain was isolated in 2012 from an outbreak of erythromelalgia-associated poxvirus in rural China in 1987 \cite{mendez2012genome}; most databases do not record this virus as zoonotic.
    \textsuperscript{[c]} Tentative serological evidence recorded \cite{kuroya_newborn_1953}.
    \textsuperscript{[d]} Serological evidence recorded \cite{maruyama_survey_1983}.
    \textsuperscript{[e]} Serological evidence first recorded from cases associated with occupational exposure \cite{centers_for_disease_control_cdc_update_1990}.
    \textsuperscript{[f]} Serological evidence recorded for Potsikum virus \cite{olaleye1990survey}, now a member of \textit{Saboya virus}. 
    \textsuperscript{[g]} Tentative evidence of viral isolation is recorded \cite{lvov_isolation_1980}.}
    \centering
    \quad \\
    {\renewcommand{\arraystretch}{1.5}
    \begin{tabular}{|l|m{0.12\linewidth}|>{\raggedright\arraybackslash}m{0.23\linewidth}|c|m{0.09\linewidth}|}
         \hline
        \textbf{Virus} & \textbf{Virus family (-viridae)} & \textbf{Animal hosts (number of spp.)} & \textbf{Prob.} & \textbf{Prior risk} \\
         \hline
         \textit{Murine respirovirus}\textsuperscript{[c]} & Paramyxo & Rodentia (9), Artiodactyla (1), Carnivora (1), Primates (1) & 0.683 & Medium \\
         \textit{Akabane orthobunyavirus}\textsuperscript{[d]} & Peribunya & Artiodactyla (31), Perissodactyla (4), Proboscidea (1) & 0.682 & High \\
         \textit{Reston ebolavirus}*$\ddag$\textsuperscript{[e]} & Filo & Chiroptera (9), Artiodactyla (1), Primates (1) & 0.680 & High \\
         \textit{Saboya virus}\textsuperscript{[f]} & Flavi & Rodentia (4), Chiroptera (1) & 0.679 & High \\
         \textit{Simian orthorubulavirus}*$\ddag$ & Paramyxo & \textit{Macaca fascicularis} (Primates) & 0.678 & High \\
         \textit{Chobar Gorge virus}*$\dagger$ & Reo & Artiodactyla (2), Chiroptera (2), Perissodactyla (1) & 0.673 & Medium \\
         Issyk-Kul virus\textsuperscript{[g]} & Nairo & Chiroptera (13) & 0.672 & --- \\
         \textit{Patois orthobunyavirus}*$\ddag$ & Peribunya & Rodentia (6), Artiodactyla (2), Didelphimorphia (2), Carnivora (1), Lagomorpha (1) & 0.667 & High \\
         \textit{Bovine fever ephemerovirus}	& Rhabdo & Artiodactyla (30), Proboscidea (1) & 0.660 & Medium \\
         \textit{Minatitlan orthobunyavirus} & Peribunya & Primates (1), Rodentia (1) & 0.654 & --- \\
         \hline
    \end{tabular}}
    \label{tab:my_label}
\end{table}

\renewcommand{\thetable}{\arabic{table}}

\end{document}